\documentclass[a4paper,fleqn,usenatbib]{mnras}

\usepackage{newtxtext,newtxmath}
\usepackage{gensymb}

\usepackage[T1]{fontenc}
\usepackage{ae,aecompl}

\usepackage{siunitx}

\usepackage{graphicx}	
\usepackage{amsmath}	
\usepackage{nccmath}
\usepackage{longtable}
\usepackage{arydshln}
\usepackage{booktabs}
\usepackage{tabularx}
\usepackage{supertabular}

\newcommand{\hmath}[1]{\texorpdfstring{{\rmfamily\textit{$#1$}}}{}}

\title[Black hole activity and spin--filament alignments]{The SAMI Galaxy Survey: impact of black hole activity on galaxy spin--filament alignments}

\author[S.\ Barsanti et al.]{Stefania Barsanti,$^{1,2,3}$\thanks{E-mail: stefania.barsanti@anu.edu.au}
Matthew Colless,$^{1,2,4}$
Francesco D'Eugenio,$^{5}$
Sree Oh,$^{6,1,2}$
\newauthor
Julia J. Bryant,$^{3,7,2}$ 
Sarah Casura,$^{8}$
Scott M. Croom,$^{3,2}$
Yifan Mai,$^{3,2}$
Andrei Ristea,$^{9,2}$
\newauthor
Jesse van de Sande,$^{3,2}$
Charlotte Welker,$^{10,11,12}$
Henry R. M. Zovaro$^{1,2}$
\vspace{0.4cm}
\\
$^{1}$Research School of Astronomy and Astrophysics, Australian National University, Canberra, ACT 2611, Australia\\
$^{2}$ARC Centre of Excellence for All Sky Astrophysics in 3 Dimensions (ASTRO 3D), Australia\\
$^{3}$Sydney Institute for Astronomy (SIfA), School of Physics, The University of Sydney, NSW 2006, Australia\\
$^{4}$Sub-Department of Astrophysics, Department of Physics, University of Oxford, Denys Wilkinson Building, Keble Rd., Oxford OX1 3RH, UK\\
$^{5}$Cavendish Laboratory and Kavli Institute for Cosmology, University of Cambridge, Madingley Rise, Cambridge, CB3 0HA, United Kingdom\\
$^{6}$Department of Astronomy, Yonsei University, Seoul 03722, Republic of Korea\\
$^{7}${Australis-USydney, School of Physics, University of Sydney, NSW 2006, Australia}\\
$^{8}$Hamburger Sternwarte, Universit\"{a}t Hamburg, Gojenbergsweg 112, 21029 Hamburg, Germany\\
$^{9}$International Centre for Radio Astronomy Research, University of Western Australia, 35 Stirling Highway, Crawley, WA 6009, Australia\\
 $^{10}$New York City College of Technology, City University of New York, 300 Jay Street, Brooklyn, NY, USA\\
 $^{11}$Centre for Computational Physics, Flatiron Institute, 160 5th avenue, New York, NY, USA\\
 $^{12}$Department of Physics \& Astronomy, Johns Hopkins University, Baltimore, MD 21218, USA\\
}

\date{Accepted 2023 September 1. Received 2023 August 16; in original form 2023 May 14}

\pubyear{2023}

\begin{document}
\label{firstpage}
\pagerange{\pageref{firstpage}--\pageref{lastpage}}
\maketitle

\begin{abstract}
The activity of central supermassive black holes might affect the alignment of galaxy spin axes with respect to the closest cosmic filaments. We exploit the SAMI Galaxy Survey to study possible relations between black hole activity and the spin--filament alignments of stars and ionised gas separately. To explore the impact of instantaneous black hole activity, active galaxies are selected according to emission-line diagnostics. Central stellar velocity dispersion ($\sigma_{\rm c}$) is used as a proxy for black hole mass and its integrated activity. We find evidence for the {\em gas} spin--filament alignments to be influenced by AGN, with Seyfert galaxies showing a stronger perpendicular alignment at fixed bulge mass with respect to galaxies where ionisation is consequence of low-ionizaition nuclear emission-line regions (LINERs) or old stellar populations (retired galaxies). On the other hand, the greater perpendicular tendency for the {\em stellar} spin--filament alignments of high-bulge mass galaxies is dominated by retired galaxies. Stellar alignments show a stronger correlation with $\sigma_{\rm c}$ compared to the gas alignments. We confirm that bulge mass ($M_{\rm bulge}$) is the primary parameter of correlation for both stellar and gas spin--filament alignments (with no residual dependency left for $\sigma_{\rm c}$), while $\sigma_{\rm c}$ is the most important property for secular star formation quenching (with no residual dependency left for $M_{\rm bulge}$). These findings indicate that $M_{\rm bulge}$ and $\sigma_{\rm c}$ are the most predictive parameters of two different galaxy evolution processes, suggesting mergers trigger spin--filament alignment flips and integrated black hole activity drives star formation quenching.
\end{abstract}

\begin{keywords}
galaxies: evolution -- galaxies: kinematics and dynamics -- galaxies: structure, fundamental parameters  -- galaxies: active -- cosmology: large-scale structure of Universe
\end{keywords}


\section{Introduction}
\label{Introduction}

The activity of central supermassive black holes in galaxies can bring striking changes in their evolution (e.g., \citealp{Cattaneo2009,Fabian2012,Heckman2014}). The role of active galactic nuclei (AGN) feedback, i.e.\ the self-regulating process that links the kinetic and thermal energy released by the AGN to the surrounding gaseous medium (see \citealp{Morganti2017} for a review), has been widely argued to be a driver of galaxy properties and quenching of star formation activity in galaxies (e.g., \citealp{Lagos2008,Dubois2016}). Star formation quenching refers to the transition of a galaxy from being star-forming to passive over time, and it is often quantified as the departure from the stellar mass--star formation rate main sequence \citep{Noeske2007,Brinchmann2004}. Moreover, the occurrence and luminosity of AGN is predicted by simulations to correlate with star formation rate, while the opposite does not stand (e.g., \citealp{Ward2022}).

In cosmological hydrodynamical simulations, the mass of the black hole has been found to be the most predictive parameter for the star-forming versus quenched status of central galaxies \citep{Piotrowska2022}. Using observations for nearby central galaxies at $z<0.034$, \citet{Terrazas2016} found that quiescent galaxies have more massive black holes than the star-forming ones at fixed stellar mass. Black hole mass is difficult to measure directly beyond the very local Universe. Since central velocity dispersion has been discovered to strongly correlate with black hole mass (e.g., \citealp{Ferrarese2000,Saglia2016}), it is often used as a proxy for this latter in large galaxy surveys. In agreement with the predictions from simulations, central stellar velocity dispersion is identified as the main galaxy property to correlate with star formation quenching \citep{Bluck2020,Brownson2022,Bluck2022} and metallicity \citep{Barone2020,Vaughan2022} in observations, especially for central galaxies and high-mass satellite galaxies. Since the mass of the black hole traces the total energy released during its growth (e.g., \citealp{Silk2012}), these results point to a scenario where it is the integrated black hole activity, with slow and steady energy injection (e.g., \citealp{Birzan2004,Hlavacek-Larrondo2012,Zinger2020}), that drives star formation quenching, rather than the instantaneous AGN activity through violent gas outflows (e.g., \citealp{Hopkins2008,Maiolino2012,Bischetti2019}).

AGN feedback might also have an impact on the intrinsic alignments of galaxies, i.e.\ the inherent correlations of galaxy shapes and spins across the large-scale structure of the Universe. According to the tidal field theory, galaxy spin is generated by torques acting on the collapsing proto-halo, which gains angular momentum from the gravitational perturbations in the tidal field \citep{Peebles1969,Porciani2002,Schafer2009,Codis2012,Codis2015}. As a consequence, one of the intrinsic alignments that we can detect is that of the galaxy spin axis with respect to the orientation of the closest cosmic filament \citep{Dubois2014,Codis2018}. Low-mass galaxies formed via gas accretion tend to have spins aligned parallel to filaments, while mergers along the filament are able to flip the spin--filament alignments of massive galaxies from parallel to perpendicular \citep{Welker2014,Welker2020}. In this context, \citet{Dubois2014} suggested that AGN feedback plays an important role in maintaining the perpendicular alignment after mergers, by preventing further gas inflow and avoiding re-alignment back to parallel. Indeed, \citet{Soussana2020} found that galaxies have more significant perpendicular trends for Horizon-AGN simulations \citep{Dubois2012} with respect to Horizon-noAGN simulations \citep{Peirani2017}. However, the difference disappears for a matched sample of galaxies, indicating that the tendency to more perpendicular alignments is due to the higher abundance of massive pressure-supported galaxies when AGN feedback is implemented. A key question that remains is whether AGN feedback is able to limit the tendency of alignments, for example via randomizing the direction of cold gas accretion early on.


Supermassive black hole masses are strongly correlated with their host bulge stellar mass (e.g., \citealp{Croton2006}). Bulge mass has been found to be a key parameter associated with both star formation quenching \citep{Lang2014,Bluck2014,Bluck2022,Dimauro2022} and galaxy spin--filament alignments \citep{Barsanti2022}. The flipping of the galaxy spin--filament alignment from parallel to perpendicular and star formation quenching can be considered as separate processes in galaxy evolution, as star formation activity is found not to have a major impact on the alignments of galaxies both in simulations \citep{Kraljic2020} and observations \citep{Barsanti2022}. The correlation of star formation rate with spin--filament alignment is due to a common dependency on bulge mass. Recently, \citet{Lagos2022} found that quenching driven by AGN feedback precedes kinematic transformation for most slow rotators in the EAGLE simulations, highlighting that the two processes take place at different times in the evolution of these galaxies.  

In this context, the following questions arise: (i)~Does black hole activity have an impact on galaxy spin--filament alignments, and, if so, is it the instantaneous or the integrated black hole activity that matters? (ii)~Are there differences in the impact of the black hole activity on stellar spin--filament alignments versus ionised gas spin--filament alignments? (iii)~What are the relative roles of central velocity dispersion and bulge mass in spin--filament alignments and star formation quenching?

We aim to address these questions by exploring whether the presence of an AGN influences the galaxy spin--filament alignment, considering stellar spins and ionised gas spins separately. We make use of the galaxy spins and cosmic filaments measured in \citet{Barsanti2022}. The spatially-resolved kinematics from the Sydney--AAO Multi-object Integral-field (SAMI) spectroscopic galaxy survey \citep{Croom2021} is used to estimate galaxy spins, while the Galaxy And Mass Assembly (GAMA) redshift survey  \citep{Driver2022} is used to reconstruct the cosmic web. In addition to active galaxies, we investigate the correlation between spin--filament alignments and central stellar velocity dispersion, taking the latter as a proxy for the black hole mass and its integrated activity. For a more complete picture, we compare the roles of central velocity dispersion and bulge mass in spin--filament alignments to star formation quenching. This analysis sheds light on the primary galaxy parameter of correlation for these galaxy evolution processes and the consequent driving physical mechanisms. To assess correlations, we implement statistical tests and the Random Forest classifier, a machine learning algorithm which has been proven to be effective in identifying causality in astronomical data \citep{Bluck2022,Piotrowska2022}.

This paper is structured as follows. We present our galaxy sample and galaxy properties in Section~\ref{Data and Galaxy Sample}. In Section~\ref{Methods}, we describe the methods used to identify galaxy spin--filament alignments and star formation quenching. In Section~\ref{Results}, we present our results about the impact of AGN on galaxy spin--filament alignments, and the roles of bulge mass and central velocity dispersion in the alignments and star formation quenching. In Section~\ref{Discussion}, we compare our findings to previous studies and we discuss their physical interpretations. Finally, we state our conclusions in Section~\ref{Summary and conclusions}. Throughout this work, we assume $\Omega_{m}=0.3$, $\Omega_{\Lambda}=0.7$ and $H_0=70$\,km\,s$^{-1}$\,Mpc$^{-1}$ as the cosmological parameters.

\section{Data and Galaxy Samples}
\label{Data and Galaxy Sample}

\subsection{The SAMI Galaxy Survey}
\label{SAMI galaxy survey}

The SAMI Galaxy Survey is a spatially-resolved spectroscopic survey of 3068 galaxies with stellar mass range $\log(M_{\star}/M_{\odot})$\,=\,8--12 and redshift range $0.004<z\leq0.115$ \citep{Bryant2015,Croom2021}. Most of the SAMI targets belong to the three equatorial fields (G09, G12 and G15) of the Galaxy And Mass Assembly survey \citep[GAMA;][]{Driver2011}. Eight massive clusters were also observed \citep{Owers2017}, but they are excluded from this work since environmental processes might affect galaxy spin--filament alignments \citep{Dubois2014} and star formation quenching \citep{Dressler1980}. The Sydney--AAO Multi-object Integral-field spectrograph (SAMI) was mounted on the 3.9\,m Anglo-Australian Telescope \citep{Croom2012}. The instrument has 13 fused optical fibre bundles (hexabundles), each containing 61 fibres of 1.6\,arcsec diameter so that each integral field unit (IFU) has a 15\,arcsec diameter \citep{Bland2011,Bryant2014}. The SAMI fibres feed the two arms of the AAOmega spectrograph \citep{Sharp2006}. The SAMI Galaxy Survey uses the 580V grating in the blue arm, giving a resolving power of $R$=1812 and wavelength coverage of 3700--5700\,\AA, and the 1000R grating in the red arm, giving a resolving power of $R$=4263 over the range 6300--7400\,\AA. The median full-width-at-half-maximum values for each arm are FWHM$_{\rm blue}$=2.65\,\AA\ and FWHM$_{\rm red}$=1.61\,\AA\ \citep{vandeSande2017}. The SAMI datacubes are characterised by a grid of 0.5$\times$0.5\,arcsec spaxels, where the blue and red spectra have pixel scales of 1.05\,\AA\ and 0.60\,\AA\ respectively. We exploit the SAMI Galaxy Survey for measurements of stellar spin axes, ionised gas spin axes, and central stellar velocity dispersions from spatially-resolved kinematics.

\subsection{The GAMA survey}
\label{GAMA galaxy survey}

The Galaxy And Mass Assembly survey (GAMA; \citealp{Driver2011,Baldry2018,Driver2022}) is a spectroscopic and photometric survey of $\sim$300,000 galaxies down to $r < 19.8$\,mag that covers $\sim$286\,deg$^{2}$ in 5 regions called G02, G09, G12, G15 and G23. The redshift range of the GAMA sample is $0<z<0.5$, with a median value of $z\sim0.25$. Most of the spectroscopic data were obtained using the AAOmega multi-object spectrograph at the Anglo-Australian Telescope, although GAMA also incorporates previous spectroscopic surveys such as SDSS \citep{York2000}, 2dFGRS \citep{Colless2001,Colless2003}, WiggleZ \citep{Drinkwater2010} and the Millennium Galaxy Catalogue \citep{Driver2005}. 

The GAMA survey’s deep and highly complete spectroscopic redshift data, combined with its wide area, high spatial resolution, and broad wavelength coverage make it an ideal galaxy sample for mapping the filaments of the cosmic web. In particular, the high spectroscopic completeness of GAMA (98.5\% in the equatorial regions; \citealp{Liske2015}) compared to the SDSS survey (> 80\%; \citealp{Driver2022}) allows to reconstruction of the filamentary structures at a smaller scale relative in comparison to the large filaments probed using SDSS in other works investigating the cosmic web (e.g., \citealp{Kraljic2021}).

\subsection{Galaxy properties}
\label{Galaxy properties}

\subsubsection{Galaxy and bulge stellar masses}
\label{Galaxy and bulge stellar masses}

Galaxy stellar masses ($M_\star$) are measured from K-corrected $g$--$i$ colours and $i$-band magnitudes for the SAMI Galaxy Survey \citep{Bryant2015} and for the whole GAMA survey \citep{Taylor2011}. \citet{Casura2022} and \citet{Barsanti2021} performed 2D photometric bulge/disc decomposition for the SAMI-GAMA galaxies and the SAMI cluster galaxies, respectively. Both photometric decompositions use the image analysis package {\sc ProFound} and the photometric galaxy profile fitting package {\sc ProFit} \citep{Robotham2017,Robotham2018}. The disc is defined as the exponential component, while the bulge corresponds to the S\'ersic component representing the light excess over the exponential component. We take advantage of the 2D bulge/disc decomposition from \citet{Casura2022}, which covers the GAMA equatorial survey regions for $z<0.1$. The decomposition is performed on the $g$-, $r$-, and $i$-band images from the Kilo-Degree Survey (KiDS; \citealp{deJong2017}). We use $r$-band bulge-to-total flux ratio (B/T) extrapolated to infinity, but the same results are found if we use the integrated quantity limited to a segment radius including the photometric fitting of the galaxy. The combination of $M_\star$ and B/T parameters allows us to explore galaxy spin--filament alignments and star formation quenching as a function of the mass of the bulge, defined as $M_{\rm bulge} \equiv M_\star\times({\rm B/T})$. In this work we use $M_{\rm bulge}$ as a tracer of mergers.

\subsubsection{Star formation proxies}
\label{Star formation proxies}

Star-formation rate (SFR) is estimated using
dust-corrected H$\alpha$ flux, assuming an intrinsic Balmer decrement H$\alpha$/H$\beta$\,=\,2.86 and the \citet{Cardelli1989} extinction law. Only star-forming spaxels are taken into account \citep{Medling2018}. We use integrated SFR values within the galaxy's semi-major axis effective radius $R_{\rm e}$. However, these measurements tend to be an underestimate of the total SFR, since they are limited to 1\,$R_{\rm e}$ aperture. Thus, we also make use of the photometric SFR catalogue assembled by \citet{Ristea2022} for the SAMI Galaxy Survey, which combined measurements obtained via the spectral energy distribution (SED)-fitting Code Investigating GALaxy Emission (CIGALE; \citealp{Noll2009}) for SAMI galaxies in the
GALEX SDSS WISE Legacy Catalog \citep{Salim2018} and the SED-fitting code
{\sc ProSpect} \citep{Robotham2020} for the remaining SAMI galaxies \citep{Bellstedt2020a,Bellstedt2020b}. Consistent results are found using the spectroscopic or the photometric SFR estimates.

\subsubsection{Diagnostics of AGN}
\label{Diagnostics of AGN}

To explore the role of the instantaneous black hole activity on galaxy spin--filament alignments we study whether the presence of an AGN influences the tendency of the alignment. We make use of emission-line diagnostics plots to identify the type of sources contributing to ionised gas. The fluxes for the H$\beta$, [O\,III] $\lambda$5007, H$\alpha$, [N\,II] $\lambda$6583, [S\,II] $\lambda$6716, and [S\,II] $\lambda$6731 emission lines are fitted with Gaussian profiles having a common width using the fitting software {\sc lzifu} \citep{Ho2016}. We measure these emission-line fluxes within $1\,R_{\rm e}$, since similar results are obtained by using the mean values of the fluxes per spaxel within $1\,R_{\rm e}$ \citep{Oh2022}. 

\subsubsection{Central velocity dispersions}
\label{Central velocity dispersions}

To investigate the role of the integrated black hole activity on galaxy spin--filament alignments and star formation quenching, we use central stellar velocity dispersion as a proxy for the black hole mass. The relation between these two parameters is the strongest among various correlations of black hole mass with other galaxy properties \citep{Piotrowska2022}. Moreover, \citet{Bluck2023} warn against the use of AGN luminosity as a proxy for the integrated black hole activity, since it is linked only to the current accretion state of the supermassive black hole.

To determine the central velocity dispersion, we exploit spatially-resolved stellar kinematics. A detailed presentation of the stellar kinematics for the SAMI Galaxy Survey, which is based on the Penalised Pixel-Fitting software (pPXF; \citealp{Cappellari2004,Cappellari2017}) with the MILES library of stellar spectra \citep{SanchezBlazquez2006,FalconBarroso2011} as templates, can be found in \citet{vandeSande2017}. $R_{\rm e}$ and ellipticities within $R_{\rm e}$ are measured using the Multi-Gaussian Expansion (MGE; \citealp{Emsellem1994,Cappellari2002}) technique. A complete description of the MGE fits for the SAMI Galaxy Survey is presented in \citet{DEugenio2021}. We estimate the central velocity dispersion ($\sigma_{\rm c}$) as the flux-weighted mean of velocity dispersions in spaxels within $0.5\,R_{\rm e}$:
\begin{equation}
  \sigma_{\rm c}=\sqrt{\frac{\sum_{i} F_{i} \sigma_{i}^{2}}{\sum_{i} F_{i}}} ~.
\end{equation}
For each galaxy we select only spaxels with stellar continuum signal-to-noise ratio (S/N)$_{i}>3$\,\AA$^{-1}$, $\sigma_{i}>35$ km$\,\rm s^{-1}$ and $\sigma_{\mathrm{err},i}<\sigma_{i}\times0.1+25$ km$\,\rm s^{-1}$, following the quality cuts of \citet{vandeSande2017}. This definition of central velocity dispersion takes into account only dispersion contributions and it does not include a rotation velocity term. Aperture correction is not applied to these measurements, since this correction changes the velocity dispersion by at most 1\% \citep{Oh2020}. The $0.5\,R_{\rm e}$ area allows us to interpret $\sigma_{\rm c}$ as a tracer of physical processes occurring in the central region of galaxies, in comparison to other galaxies parameters, such as $M_{\star}$ and $M_{\rm bulge}$, which are measured on the whole galaxy spatial scales. 

Some previous works, such as \citet{Bluck2022}, use fibre measurements of velocity dispersions in addition to those derived from spatially-resolved spectroscopy. For a consistent comparison with these previous studies, we take advantage of central velocity dispersions ($\sigma_{\star}$) estimated from fibre spectra for GAMA DR4 \citep{Driver2022}. The description of how these measurements are obtained can be found in Appendix~\ref{Central velocity dispersions from GAMA spectra}. In this Appendix, we also compare $\sigma_{\star}$ to $\sigma_{\rm c}$, concluding that our results do not change if we use one instead of the other. The same findings stand for SAMI velocity dispersions estimated within a central 3\,arcsec aperture, in agreement with the comparison between this latter estimates and $\sigma_{\star}$ from \citet{Scott2017}.

\subsection{Galaxy samples}
\label{Galaxy samples}

In this work we make use of the SAMI galaxy sample selected by \citet{Barsanti2022} to study spin--filament alignments, which is limited to $9<\log(M_\star/M_{\odot})<12$ to ensure reliable stellar kinematic velocity maps \citep{vandeSande2017}. The galaxies have measured ellipticities and kinematic position angles (PA) with 1$\sigma$ uncertainty $\delta{\rm PA}<25$\degree. Their final SAMI galaxy sample comprises 1121 galaxies with measured $M_\star$ and B/T. 

\begin{figure*}
\includegraphics[scale=0.285]{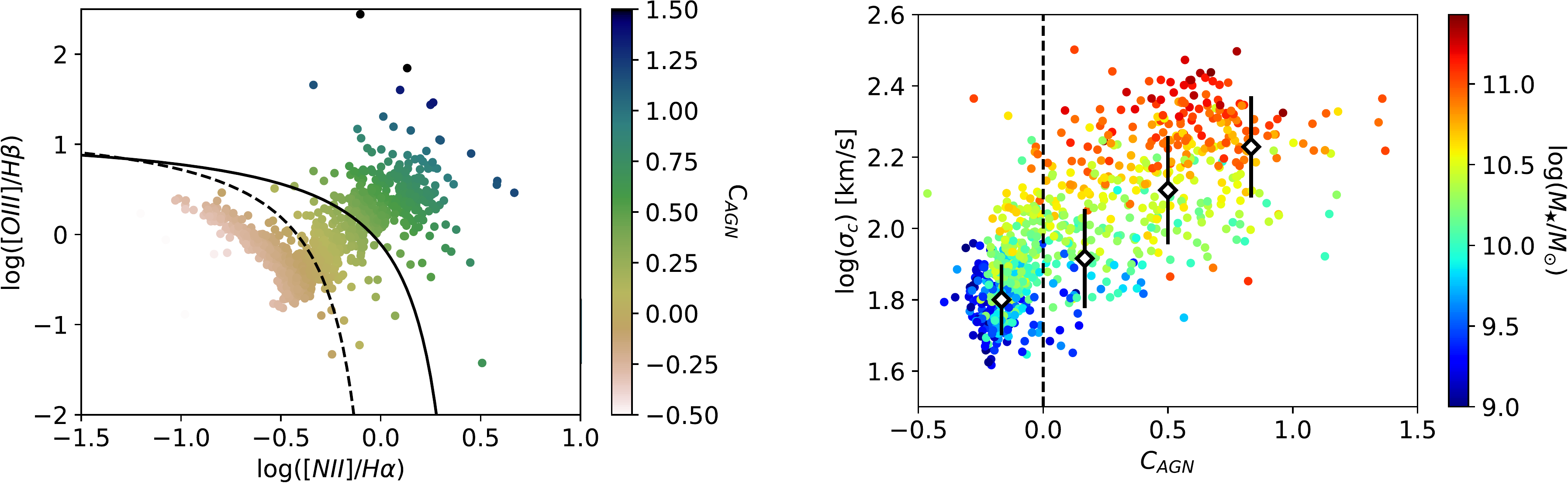}
\caption{Left panel: BPT diagram for Sample~A (977 SAMI galaxies). Galaxies are colour-coded according to $\rm C_{\rm AGN}$, the orthogonal departure in log scale from the star-forming limit of \citet{Kauffmann2003} (dashed line). The solid line represents the theoretical maximal star-forming limit of \citet{Kewley2001}. Right panel: relation between $\rm C_{\rm AGN}$ (proxy for instantaneous black hole activity) and $\sigma_{\rm c}$ (proxy for integrated black hole activity). Points are colour-coded by stellar mass. The dashed line marks the star-forming limit $\rm C_{\rm AGN}=0$. The black diamonds show running medians with root-mean square dispersion, highlighting a strong positive monotonic relationship between the two quantities. }
\label{cAGN_diagrams}
\end{figure*}

To understand the impact of AGN on galaxy spin--filament alignments, from the sample of \citet{Barsanti2022} characterised by 1121 galaxies we select 977 galaxies (87\%) with measured Baldwin, Phillips \& Terlevich (BPT) emission-line diagnostic lines \citep{Baldwin1981,Veilleux1987}. A signal-to-noise limit higher than 3\,\AA$^{-1}$ for each emission line is also used. From here on, we refer to this sample as Sample~A. The BPT diagram allows us to separate star-forming galaxies from those dominated by an AGN \citep{Kewley2001}. The left panel of Figure~\ref{cAGN_diagrams} shows the BPT diagram for Sample~A using the H$\alpha$, H$\beta$, [N\,II], and [O\,III] line flux ratios. Following \citet{Oh2022}, we define the `AGN contribution', $\rm C_{\rm AGN}$, as the orthogonal departure in log scale from the star-forming limit line of \citet{Kauffmann2003}. By definition, low and high values of $\rm C_{\rm AGN}$ correspond to star-forming galaxies and AGN respectively. 

All 977 galaxies in Sample~A have reliable measurements of central velocity dispersion $\sigma_{\rm c}$, with most of the galaxies ($\sim$83\%) having more than 90\% of spaxels within $0.5\,R_{\rm e}$ fulfilling the quality cuts listed in Section~\ref{Central velocity dispersions}. The right panel of Figure~\ref{cAGN_diagrams} shows the relation between $\rm C_{\rm AGN}$ (proxy for instantaneous black hole activity) and $\sigma_{\rm c}$ (proxy for integrated black hole activity). The two quantities have a strong positive monotonic relationship (Spearman rank correlation test: $\rho=0.74$, $p_{\rm S}=10^{-139}$), with an average root-mean square dispersion of only 0.14 dex. Points are colour-coded by stellar mass, showing that both $\rm C_{\rm AGN}$ and $\sigma_{\rm c}$ correlate with $\log(M_{\star}/M_{\odot})$. The scatter in $\sigma_{\rm c}$ values increases for higher $\rm C_{\rm AGN}$: this underlies how star formation is a smooth and regular process, while AGN activity is bursty and irregular. We note that $\rm C_{\rm AGN}$ has limitations as AGN identifier, since it is not able to take into account other ionisation sources, such as old stellar populations. Thus, in Section~\ref{Results} we also investigate active galaxies using the WHAN diagram \citep{CidFernandes2011}. In addition, we separate galaxies according to [S\,II]-based diagnostics \citep{Kewley2006}, which better distinguish Seyfert galaxies having a supermassive black hole from low-ionisation nuclear emission-line regions (LINERs; \citealp{Heckman1980}).

To address the role of integrated black hole activity in spin--filament alignments and star formation quenching, we select from Sample~A those galaxies that have measurements of star formation rates. This allows us to study these galaxy evolution processes consistently for 845 SAMI galaxies, hereafter referred to as Sample~B. 

\section{Methods}
\label{Methods}

\subsection{Stellar spin--filament alignments}
\label{Stellar spin--filament alginments}

We use the galaxy spin--filament alignments measured by \citet{Barsanti2022}. In summary, they take advantage of spatially-resolved stellar kinematics from the SAMI Galaxy Survey to compute kinematic position angles ($\rm PA_{stars}$), using the {\sc fit\_kinematic\_pa} routine (see Appendix~C of \citealp{Krajnovic2006}). The orientation of the galaxy spin axes is identified following the 3D thin-disc approximation \citep{Lee2007,Kraljic2021,Tudorache2022}. This method requires position angles and inclination angles. These latter are estimated according to \citet{Haynes1984}, where the intrinsic flatness parameter corresponds to the mean value 0.171 of the selected SAMI sample and the galaxy axial ratio is measured from the ellipticity computed within $R_{\rm e}$.

The reconstruction of the cosmic filaments is carried out implementing the {\sc DisPerSe} structure extractor \citep{Sousbie2011a,Sousbie2011b}. {\sc DisPerSe} is a parameter-free and scale-free topologically motivated algorithm, which identifies 3D structures of the cosmic web starting from a point-like distribution, without making any assumption about its geometry or homogeneity. For the galaxy distribution, we use 35,882 GAMA galaxies selected within the SAMI redshift range ($0<z<0.13$), running {\sc DisPerSe} with a 3$\sigma$ persistence threshold. 3D models of the density fields can be found at \href{https://skfb.ly/o9MXv}{this URL}, while models of the cosmic filaments are displayed at \href{https://skfb.ly/o9MXz}{this URL}. Figure~3 of \citet{Barsanti2022} shows the projected network of filaments, the GAMA galaxies used to reconstruct the cosmic web, and the SAMI galaxies with measured spin. 

Finally, each SAMI galaxy is assigned to the closest filament using the 3D Euclidean distance. Following \citet{Welker2020}, the `Finger of God' effect due to the spurious elongation of galaxy groups and clusters in redshift space \citep{Jackson1972} is taken into account by reassigning SAMI galaxies initially assigned to filaments with |$\cos\alpha$|\,>\,0.9 to the closest filament with |$\cos\alpha$|\,<\,0.9, where $\alpha$ is the angle between the filament and the GAMA line-of-sight.

The galaxy spin--filament alignment is measured as the absolute value of the cosine of the angle between the galaxy spin axis and the orientation vector of the closest filament in 3D Cartesian coordinates, $|\cos\gamma|$. This varies over the range [0,1], with |$\cos\gamma$|=1 meaning the galaxy spin axis is parallel to the filament while |$\cos\gamma$|=0 means the galaxy spin axis is perpendicular to the filament. A schematic representation of these cosmic web metrics can be found in Figure~5 of \citet{Barsanti2022}. Here, stellar spin--filament alignments are represented as |$\cos\gamma$|$_{\rm stars}$.

\subsection{Gas spin--filament alignments}
\label{Gas spin--filament alignments}

Taking advantage of the SAMI spatially-resolved H$\alpha$ velocity maps, we also explore the impact of AGN on gas spin--filament alignments, where the orientation of the galaxy spin axes are identified using the gas kinematic position angles ($\rm PA_{gas}$). We use the gas spin--filament alignments estimated in \citet{Barsanti2022} and represented here as |$\cos\gamma$|$_{\rm gas}$. 

\begin{figure}
\includegraphics[width=\columnwidth]{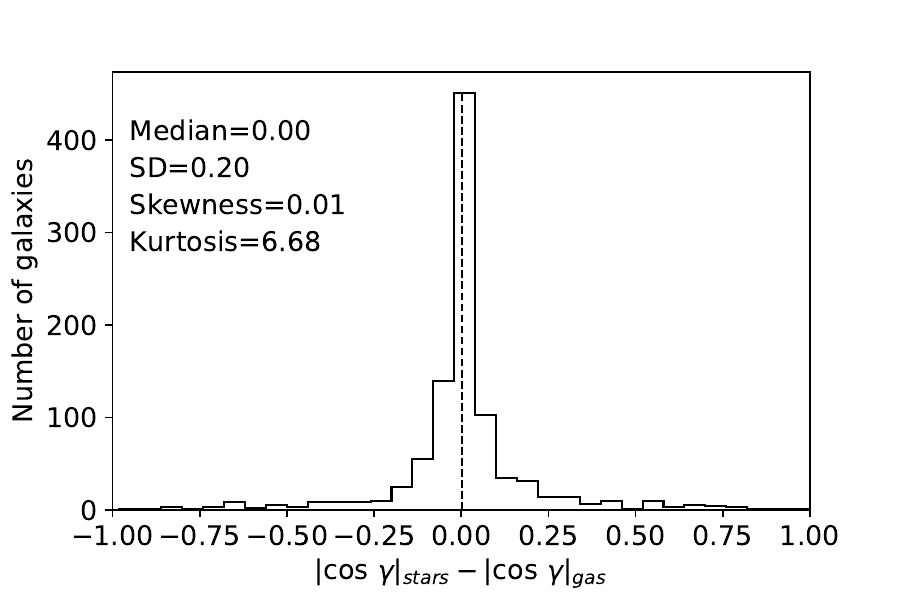}
\caption{Distribution of the differences between stellar and gas spin--filament alignments for Sample~A (977 SAMI galaxies). The black dashed line represents the median$\sim$0 dex. The high kurtosis indicates heavy tails and outliers: $\sim$15\% of galaxies have |$\cos\gamma|_{\rm stars}$ different from $|\cos\gamma|_{\rm gas}$ at >1\,SD.}
\label{alignemnts_gas_stars}
\end{figure}

The comparison between the stellar spin--filament alignments and the gas alignments for the 977 SAMI galaxies of Sample~A (defined in Section~\ref{Galaxy samples}) is shown in Figure~\ref{alignemnts_gas_stars}. The distribution of |$\cos\gamma|_{\rm stars}-|\cos\gamma|_{\rm gas}$ has median $\sim0$, standard deviation SD $\sim0.2$, skewness $\sim0.01$ and a kurtosis $\sim7$ (for reference, the normal distribution has zero valued skewness and kurtosis). The skewness $\sim0$ highlights the symmetry of the distribution, suggesting no particular shift towards more parallel or more perpendicular tendencies for stars or gas. The high kurtosis indicates heavy tails and outliers: $\sim$15\% of galaxies have |$\cos\gamma|_{\rm stars}$ different from $|\cos\gamma|_{\rm gas}$ at >1\,SD. Since the signal that we are studying is expected to be weak at low redshift ($z<0.1$; \citealp{Codis2018}), it is interesting to explore whether the black hole activity impacts differently |$\cos\gamma|_{\rm stars}$ and $|\cos\gamma|_{\rm gas}$.


\subsection{Star formation quenching}
\label{Star formation quenching}

To better understand the roles of bulge mass and central velocity dispersion, and hence the physical mechanisms that these parameters trace in galaxy evolution, we explore these galaxy properties not only for galaxy spin--filament alignments but also for star formation quenching. We quantify star formation quenching as the departure from the main sequence in the stellar mass--star formation rate plane, in agreement with previous studies, such as \citet{Bluck2014,Bluck2016,Bluck2020,Bluck2022} and \citet{Brownson2022}.  

\begin{figure}
\includegraphics[scale=0.265]{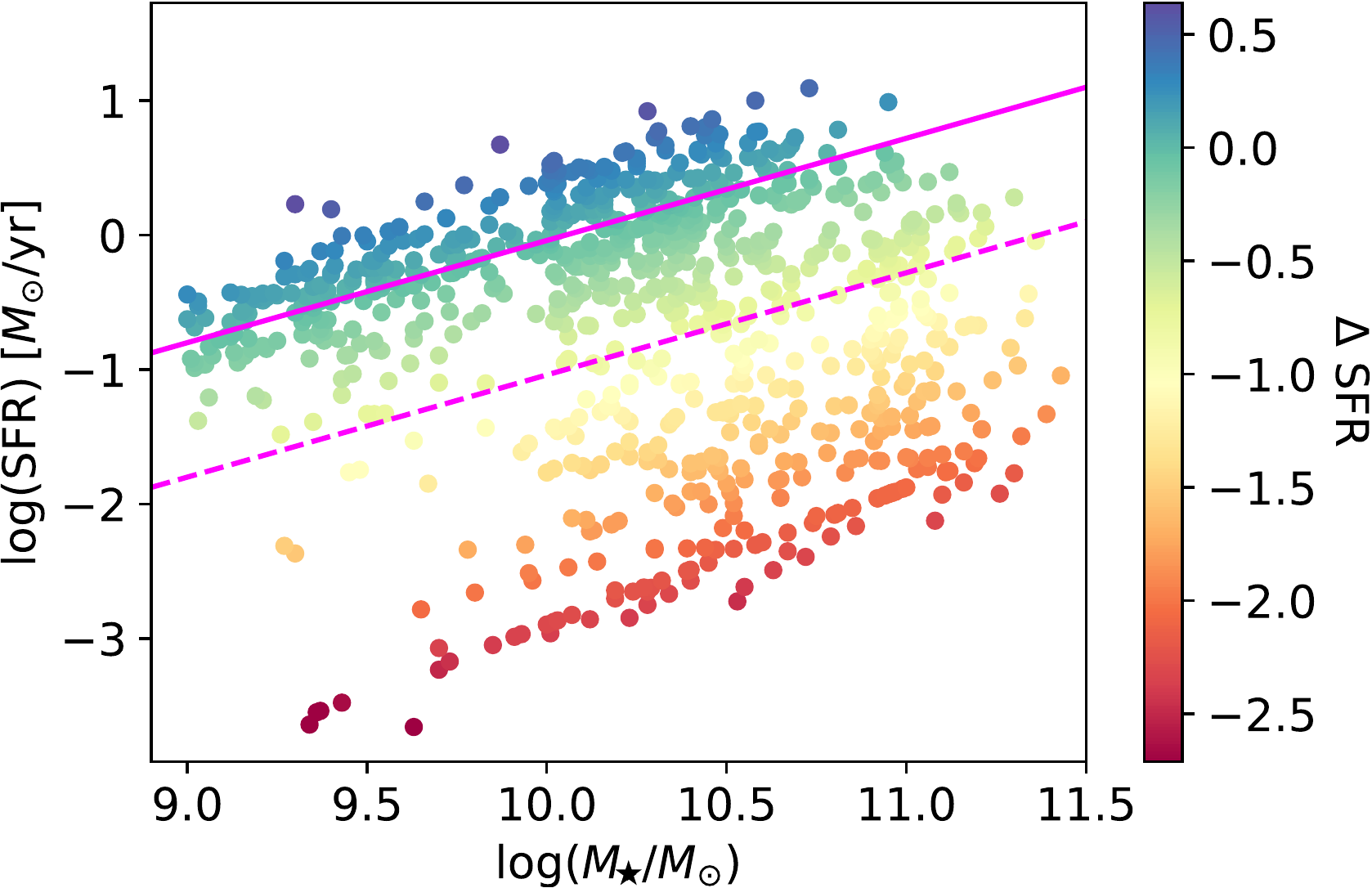}
\caption{Global $M_{\star}$--SFR plane for Sample~B (845 SAMI galaxies). Galaxies are colour-coded by their logarithmic offset in star formation rate from the main sequence fit of \citet[][solid magenta line]{Renzini2015}; the dashed magenta line lies 1\,dex below the main sequence and approximately separates star-forming and quenched galaxies.}
\label{SFR_plots}
\end{figure}

Figure~\ref{SFR_plots} displays the global $M_{\star}$--SFR plane for the 845 SAMI galaxies of Sample~B (defined in Section~\ref{Galaxy samples}). We plot SED-fitting SFRs (see Section~\ref{Star formation proxies}). Galaxies are colour-coded according to $\Delta$(SFR), the logarithmic offset in star formation rate from the main sequence fit of \citet[][solid magenta line]{Renzini2015}. $\Delta$(SFR)=0 marks galaxies on the main sequence, while $\Delta$(SFR)>0 or $\Delta$(SFR)<0 values represent enhanced or quenched star formation with respect to the main sequence, respectively. The value $\Delta$(SFR$)=-1$ (dashed magenta line) separates the star forming and quenched galaxies, and is often used to represent green valley galaxies (e.g., \citealp{Brownson2022}). 

In this work we identify passive galaxies at $\Delta$(SFR$)<-1$ and star-forming galaxies at $\Delta$(SFR$)>-0.5$, similarly to \citet{Bluck2020,Bluck2022}. Our results do not change if we shift the cut limits for $\Delta$(SFR$)$ by 0.5 dex, we use spectroscopic SFRs integrated within 1$\,R_{e}$ which are underestimated by $\sim$0.3 dex or if we divide galaxies according to specific star formation rate at $\log({\rm sSFR/yr^{-1}})=-11$ following \citet{Piotrowska2022}.

\subsection{Correlation and causality tests}
\label{Correlation and causality tests}

In this work we explore several statistical tests and machine learning algorithms to identify correlation and causality. To assess the correlation between two variables and its statistical significance, we make use of the Spearman rank correlation test (correlation coefficient, $\rho$, and $p$-value, $p_{\rm S}$). We also estimate partial correlation coefficients, which enable us to explore the true correlation between two parameters while controlling for a third quantity, avoiding the cross-correlation driven by their dependency on the third property \citep{Lawrance1976,Bait2017,Bluck2019,Baker2022}. To investigate whether a distribution is significantly different from the null hypothesis or with respect to another distribution, we use the Kolmogorov-Smirnov test (K-S test; \citealp{Lederman1984}; $p_{\rm K-S}$). $p$-values<0.05 are taken to indicate statistically significant results.

To understand which galaxy parameters primarily correlate with spin--filament alignments and star formation quenching, we apply the partial least squares regression technique (PLS; \citealp{Wold1966,Hoskuldsson1988}). The PLS technique allows us to estimate the contribution to the variance of the parameter of interest by each other parameter. We use the {\sc PLSRegression} python function from the open source machine learning package {\sc scikit-learn} for python \citep{Pedregosa2012}, following the approach described in \citet{Oh2022} and \citet{Barsanti2022}. 

Finally, we explore the Random Forest classification technique, a supervised machine learning algorithm, which allows us to estimate the relative importance of the various galaxy properties for predicting the tendency of the galaxy spin--filament alignment and the star formation state. \citet{Piotrowska2022} and \citet{Bluck2022,Bluck2023} show that this method is highly effective in finding causality within complex inter-correlated astronomical data and that it is more robust than conventional correlation-based approaches. In a nutshell, a Random Forest is a combination of multiple decision trees based on bootstrapped
random sampling. The whole sample is separated in half into the training and testing sets. At each decision split, the algorithm uses the galaxy properties to predict a given class of the training set (in this work star-forming versus passive classification, or parallel alignment versus perpendicular alignment classification; see Appendix~\ref{Examples of decision trees from Random Forest analysis} for the related examples of decision trees). To take into account the probability of incorrect classification, the Gini impurity (see Section~3.2 of \citealp{Bluck2023} for a summary of the formulae) is estimated at each split. The optimal galaxy parameter and its value are chosen to minimise the Gini impurity. We make use of the Gini technique to assess feature importance. To assure no bias in the the estimates of feature importances, we follow \citet{Bluck2022} giving the parameters in logarithmic units, subtracting the median
value, and normalising by the interquartile range.

The findings of this work do not change if we apply the permutation method, another calculation technique for estimating feature importance. The permutation technique is based on randomly shuffling the feature values, breaking the relationship between the features and the classification, thus a drop in the algorithm accuracy is indicative of how much the classification depends on a feature. The same conclusion about the consistency between the two feature importance techniques is found by \citet{Piotrowska2022}. After the training of the classifier, it possible to extract the relative importance of each explored parameter for solving the classification problem, averaging over all the implemented trees.

Following \citet{Piotrowska2022}, we implement the Random Forest algorithm using the {\sc RandomForestClassifier} class from the python {\sc scikit-learn} package. We set the number of decision trees {\fontfamily{qcr}\selectfont n\_estimators} to be 150 and the number of input features to consider at each split {\fontfamily{qcr}\selectfont max\_features} value to {\sc "sqrt"}. The {\fontfamily{qcr}\selectfont min\_samples\_leaf} parameter, i.e.\ the minimum number of objects required to be at node (default value is 1), is optimised to maximise the Area Under the Curve (AUC) of the True Positives Rate versus the False Positives Rate plot \citep{Fawcett2006,Bluck2022,Piotrowska2022}. AUC is a measure of algorithm performance, ranging from 0.5 for a random guess to 1.0 for a perfect classifier. To ensure that there are no biases in the estimates

\section{Results}
\label{Results}

Our aim is to understand whether black hole activity, either instantaneous or integrated, has an impact on galaxy spin--filament alignments. We explore whether the presence of an AGN influences the tendency of the alignments, studying separately stellar spins and ionised gas spins for active galaxies. In order to take into account the integrated black hole activity, we investigate the role of central velocity dispersion as a black hole mass proxy and its importance relative to bulge mass. In fact, these two parameters have been found to be strongly linked to galaxy evolution processes such as spin--filament alignments (e.g., \citealp{Barsanti2022}) and secular star formation quenching (e.g., \citealp{Bluck2022}).

\subsection{Impact of AGN on spin--filament alignments}
\label{Impact of AGN}

\begin{figure}
\includegraphics[width=\columnwidth]{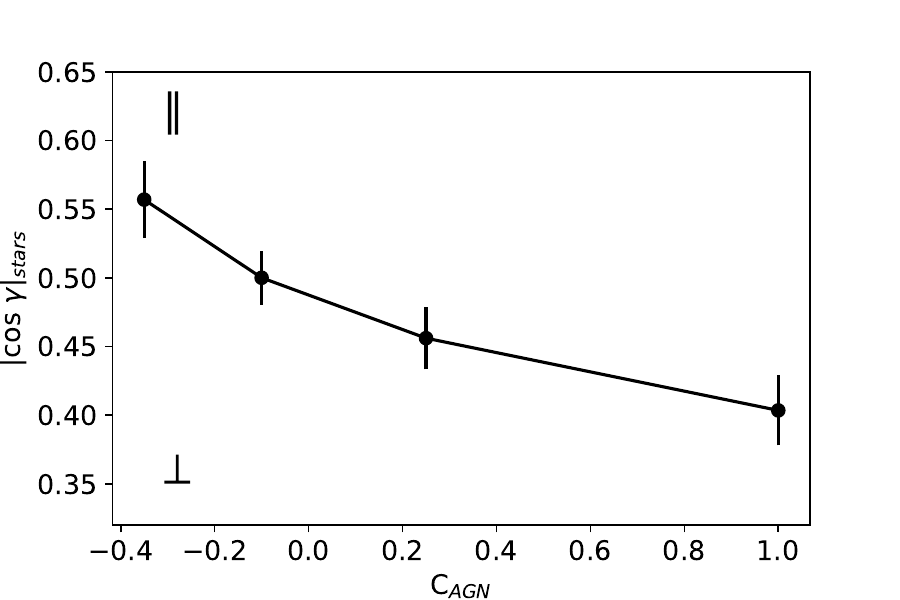}
\includegraphics[width=\columnwidth]{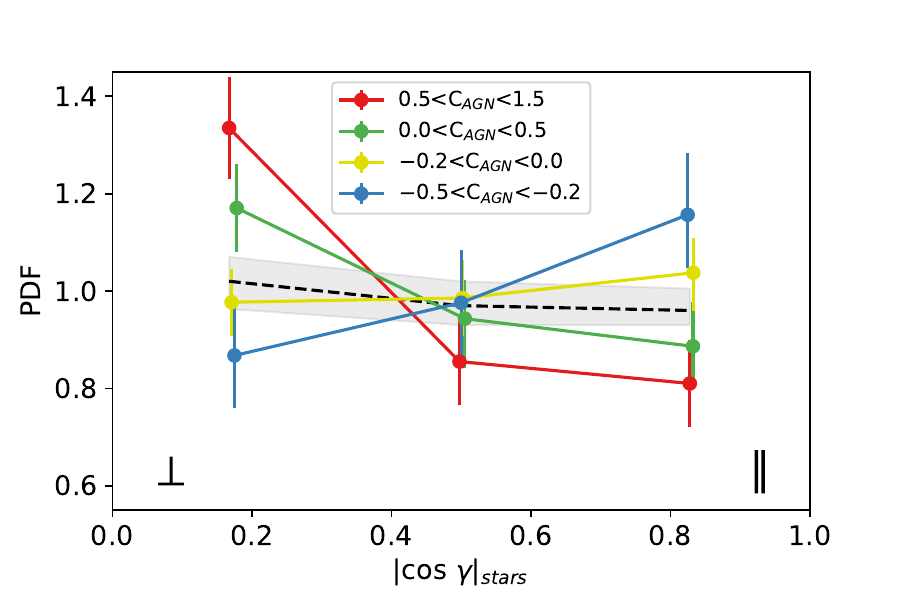}
\includegraphics[width=\columnwidth]{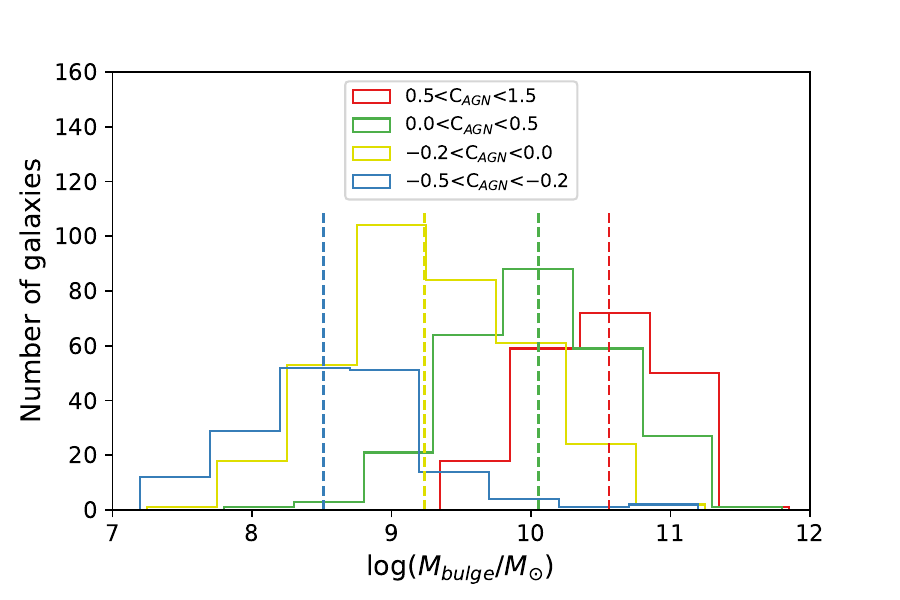}
\caption{Top: Mean $|\cos\gamma|_{\rm stars}$ values for bins in C$_{\rm AGN}$, using Sample~A (977 SAMI galaxies), showing standard errors on the mean. The mean of $|\cos\gamma|_{\rm stars}$ decreases with increasing C$_{\rm AGN}$, indicating a relative shift from parallel to perpendicular spin--filament alignments. Middle: PDFs of $|\cos\gamma|_{\rm stars}$ for four bins in C$_{\rm AGN}$. We show bootstrap error bars from 1000 random samples. The black dashed line and the shaded region represent the reconstructed
uniform distribution with its 95\% confidence intervals. The PDF is skewed towards more parallel spin--filament alignments for low values of C$_{\rm AGN}$ and to more perpendicular alignments for higher values of C$_{\rm AGN}$. Bottom: Histograms for $\log(M_{\rm bulge}/M_{\odot})$, with dotted lines showing the median value for each galaxy subsample. To increase the visibility of the distributions we plot a slightly different x-baseline for each subsample. Galaxies with higher values of C$_{\rm AGN}$ have more massive bulges, while those with lower C$_{\rm AGN}$ have lower $M_{\rm bulge}$.}
\label{cAGN_results}
\end{figure}

\subsubsection{C$_{\rm AGN}$ is a secondary tracer with respect to $M_{\rm bulge}$}
\label{cAGN is a secondary tracer}

The instantaneous activity of AGN might have an impact on galaxy spin--filament alignments. For Sample~A (977 SAMI galaxies) we study whether the position of a galaxy in the BPT diagram correlates with the angle between the galaxy stellar spin axis and the orientation vector of the closest filament, i.e.\ $|\cos\gamma|_{\rm stars}$ (see Section~\ref{Stellar spin--filament alginments}). The top panel of Figure~\ref{cAGN_results} shows the average $|\cos\gamma|_{\rm stars}$ values as a function of the AGN contribution C$_{\rm AGN}$ (see left panel of Figure~\ref{cAGN_diagrams}). The mean $|\cos\gamma|_{\rm stars}$ value decreases with increasing C$_{\rm AGN}$, indicating a relative shift from parallel to perpendicular spin--filament alignments as this quantity increases. Applying the Spearman test, we find $\rho=-0.090$ and $p_{\rm S}=0.005$, which highlights a significant correlation between $|\cos\gamma|_{\rm stars}$ and C$_{\rm AGN}$. 

In the middle panel of Figure~\ref{cAGN_results} we study the probability distribution function (PDF) of $|\cos\gamma|_{\rm stars}$ for four ranges in C$_{\rm AGN}$. To assess the statistical significance of each trend, we apply the K-S test to test the null hypothesis that $|\cos\gamma|_{\rm stars}$ has a uniform distribution. To account for possible observational bias, we compare each distribution to the null hypothesis reconstructed by generating 3000 randomised samples where the galaxy spins are fixed, but the galaxy positions are shuffled \citep{Tempel2013a,Tempel2013b,Kraljic2021,Barsanti2022}. The results from the K-S test are reported in Table~\ref{KS_Results}. For lower values of C$_{\rm AGN}$ (i.e.\ more star-forming galaxies), the PDF is skewed towards more parallel spin--filament alignments (i.e.\ higher values of $|\cos\gamma|_{\rm stars}$), while for higher values of C$_{\rm AGN}$ (i.e.\ more AGN-like galaxies) the tendency is towards more perpendicular alignments. The K-S test confirms significant results for the lowest and highest values of C$_{\rm AGN}$ detecting $p_{\rm K-S}=0.048$ and $p_{\rm K-S}=0.004$, respectively. 

\begin{table*}
  \caption{Galaxy spin--filament alignments. Column~1 lists the galaxy property or type of AGN classification, column~2 the set of galaxy subsamples for that property/classification, column~3 the number of galaxies in each subsample, column~4 the type of galaxy spin, column~5 the average |$\cos\gamma$|, column~6 the $p$-value from the K-S test with respect to the reconstructed uniform distribution, and column~7 the tendency of the alignment. Significant $p$-values (<0.05) are highlighted in bold.}
\makebox[\textwidth][c]{
    \begin{tabular}{lcccccc}
\toprule
  Galaxy Property  & Selection & $N_{\rm gal}$& Galaxy spin & <|$\cos\gamma$|> & $p_{\rm K-S}$ & Alignment \\
 \midrule
C$_{\rm AGN}$ & [$-$0.5; $-$0.2] & 166 &  stars & 0.577$\pm$0.022 & \textbf{0.048} & $\parallel$ \\
 & [$-$0.2; 0.0] & 347 & &0.500$\pm$0.016  & 0.838 & \\
 & [0.0; 0.5] & 264 & &0.446$\pm$0.018& 0.281 &   \\
 & [0.5; 1.5] & 200 & &0.403$\pm$0.020& \textbf{0.004}& $\perp$\\
\midrule
WHAN classification \& & strong AGN & 94 & stars &  0.442$\pm$0.025 & 0.094 & \\
$10<\log{(M_{\rm bulge}/M_{\odot})}<12$& weak AGN & 63 & & 0.512$\pm$0.037  & 0.197 & \\
 & retired  & 86 & & 0.355$\pm$0.029& \textbf{0.009} & $\perp$  \\
\midrule
WHAN classification  \& & strong AGN & 94 & gas &  0.325$\pm$0.024 & \textbf{0.006} & $\perp$ \\
$10<\log{(M_{\rm bulge}/M_{\odot})}<12$  & weak AGN & 63 & & 0.344$\pm$0.034  & \textbf{0.005} & $\perp$ \\
  & retired & 86 & & 0.441$\pm$0.029& 0.096 &  \\
\midrule
$\rm [S\,II]$-based classification  \& & AGN & 190 & gas &  0.395$\pm$0.019 & \textbf{0.006} & $\perp$  \\
$10<\log{(M_{\rm bulge}/M_{\odot})}<12$  & LINER & 110 & & 0.447$\pm$0.023  & 0.238 & \\
 & Seyfert & 80 & & 0.370$\pm$0.028& \textbf{0.011} & $\perp$  \\
 & star-forming & 224 & & 0.422$\pm$0.018& \textbf{0.020} & $\perp$  \\
 \midrule
X-ray classification in G09 \&  & All galaxies & 131 & gas &  0.402$\pm$0.024 & \textbf{0.024} & $\perp$  \\
 $10<\log{(M_{\rm bulge}/M_{\odot})}<12$ & X-ray AGN & 48 & & 0.387$\pm$0.038  & \textbf{0.012} & $\perp$ \\
  & no X-ray AGN & 83 & & 0.452$\pm$0.032& \textbf{0.040} & $\perp$  \\
\bottomrule
\end{tabular}
}
\label{KS_Results}
\end{table*}

We find C$_{\rm AGN}$ strongly correlates with $M_{\rm bulge}$ according to the Spearman test, with $\rho=-0.783$ and $p_{\rm S}=10^{-202}$. The bottom panel of Figure~\ref{cAGN_results} shows the histograms as a function of $M_{\rm bulge}$ for the four ranges in C$_{\rm AGN}$. Galaxies with higher values of C$_{\rm AGN}$ have more massive bulges, while those with lower C$_{\rm AGN}$ are characterised by low $M_{\rm bulge}$ values. This is in agreement with the expected $M_{\rm bulge}$ dependency of the spin--filament alignments found by \citet{Barsanti2022}. 

\begin{figure}
\includegraphics[scale=0.27]{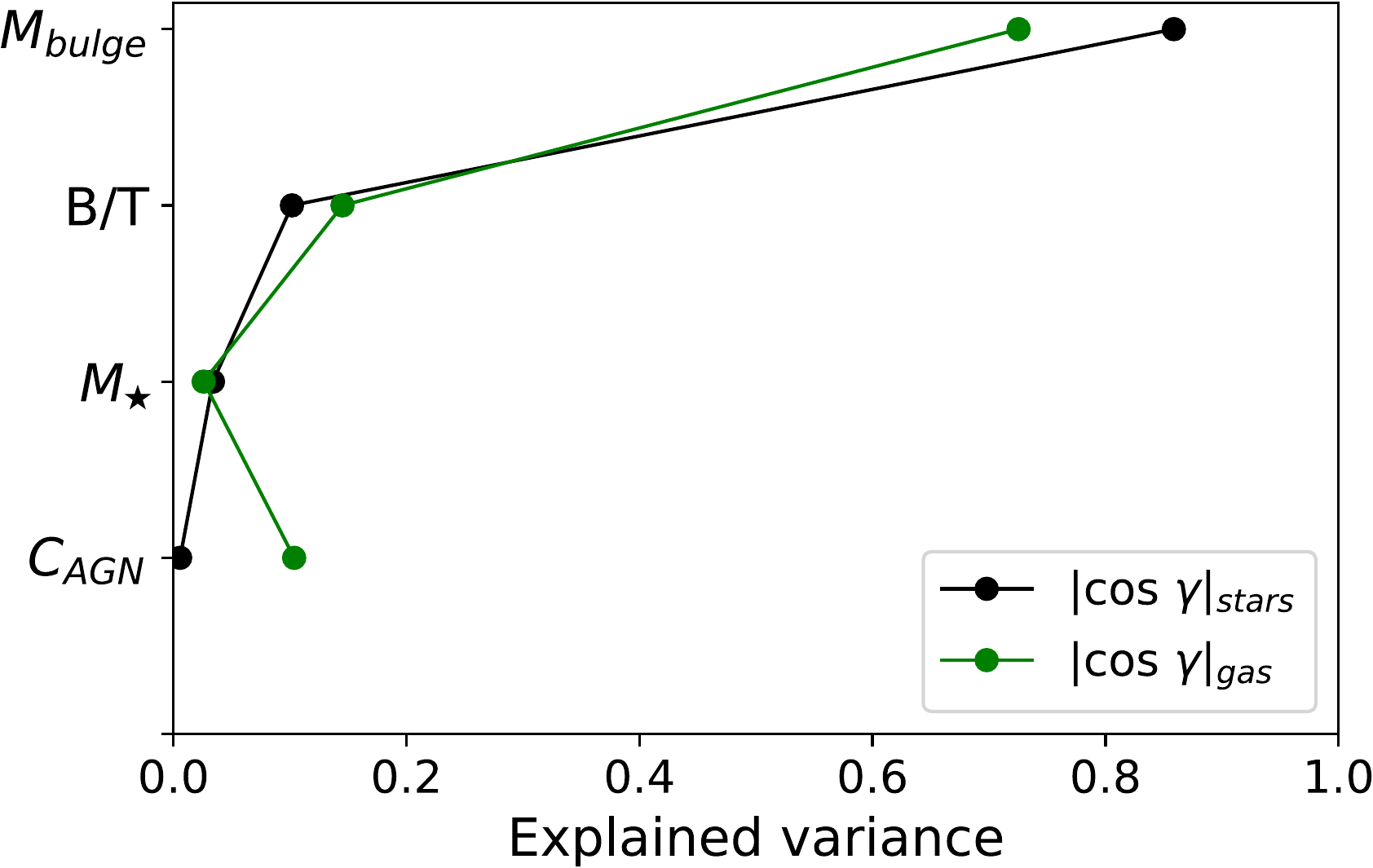}
\caption{Fractions of the variance in $|\cos\gamma|_{\rm stars}$ and $|\cos\gamma|_{\rm gas}$ explained by different galaxy properties according to the PLS regression method. Most of the variance is explained by $M_{\rm bulge}$. A $\sim$10\% higher contribution from C$_{\rm AGN}$ is found for $|\cos\gamma|_{\rm gas}$ with respect to $|\cos\gamma|_{\rm stars}$.}
\label{PLStechnique_cAGN_fig}
\end{figure}

To assess whether C$_{\rm AGN}$ or $M_{\rm bulge}$ is the primary parameter of correlation, we apply the PLS technique, investigating bulge mass, stellar mass, bulge-to-total flux ratio and C$_{\rm AGN}$. We find that 86\% of the variance in $|\cos\gamma|_{\rm stars}$ can be explained by $M_{\rm bulge}$, followed by B/T and $M_{\star}$, in agreement with the right panel of Figure~8 from \citet{Barsanti2022}. Only 0.6\% is explained by C$_{\rm AGN}$. Thus, C$_{\rm AGN}$ shows a correlation with spin--filament alignments due to its dependency on $M_{\rm bulge}$, and it is a secondary tracer with respect to $M_{\rm bulge}$. For gas spin--filament alignments (see Section~\ref{Gas spin--filament alignments}), we find that the fraction of the variance explained by $M_{\rm bulge}$ decreases to 73\% while the fraction explained by C$_{\rm AGN}$ increases to 10\%. These results are reported in Table~\ref{PLStechnique_cAGN_table} and Figure~\ref{PLStechnique_cAGN_fig}, pointing to a stronger role for C$_{\rm AGN}$ in ionised gas spin--filament alignments relative to stellar alignments, although still secondary to $M_{\rm bulge}$.

\begin{table}
\centering
\caption{Partial least squares (PLS) regression for stellar and gas spin--filament alignments. Column~1 lists the analysed galaxy property, columns~2 and~3 the explained variance in $|\cos\gamma|_{\rm stars}$ and $|\cos\gamma|_{\rm gas}$, respectively.}
\label{PLStechnique_cAGN_table}
\begin{tabular}{@{}lSS@{}}
\toprule
Galaxy property & {Variance in $|\cos\gamma|_{\rm stars}$} & {Variance in $|\cos\gamma|_{\rm gas}$} \\
 & {(\%)} & {(\%)} \\
\midrule
$\log(M_{\rm bulge}/M_{\odot})$ & 85.9 & 72.5 \\
B/T & 10.2& 14.5 \\
$\log(M_\star/M_{\odot})$ & 3.4& 2.6 \\   
$\rm C_{AGN}$ & 0.6& 10.4 \\
\bottomrule
\end{tabular}
\end{table}

\subsubsection{Link between AGN and ionised gas spin--filament alignments}
\label{AGN influence gas}

Since C$_{\rm AGN}$ is a secondary tracer of spin--filament alignments with respect to $M_{\rm bulge}$, we choose to study galaxies from Sample~A in two bulge mass ranges: $8<\log{(M_{\rm bulge}/M_{\odot})}<10$ and $10<\log{(M_{\rm bulge}/M_{\odot})}<12$. The analysis at low $M_{\rm bulge}$ range might reveal whether AGN feedback is able to limit the tendency of the alignments. The high $M_{\rm bulge}$ range allows us to investigate whether a more perpendicular tendency is seen for AGN relative to galaxies that have stopped forming stars, which might imply a possible role for AGN feedback in maintaining the perpendicular alignment for galaxies with massive bulges. 

Although C$_{\rm AGN}$ is a useful parameter for detecting quantitative correlations and trends with $|\cos\gamma|$, it does not take into account other ionisation sources, such as old stellar populations, and more care needs to be taken in the classification of AGN-like galaxies. Thus, we make use of the WHAN diagram \citep{CidFernandes2011} to select various subclasses of AGN-like galaxies: {\em retired} galaxies (i.e.\ galaxies that have stopped star formation activity and whose ionisation is due to old stellar populations), which have  equivalent widths EW(H$\alpha$)<3\,\AA; {\em weak} AGN with 3\,\AA\,<\,EW(H$\alpha$)\,<6\,\AA\ and log([N\,II]/H$\alpha$)\,>\,$-0.4$; and {\em strong} AGN with EW(H$\alpha$)\,>\,6\,\AA\ and log([N\,II]/H$\alpha$)\,>\,$-0.4$. 

\begin{figure*}
\includegraphics[width=\columnwidth]{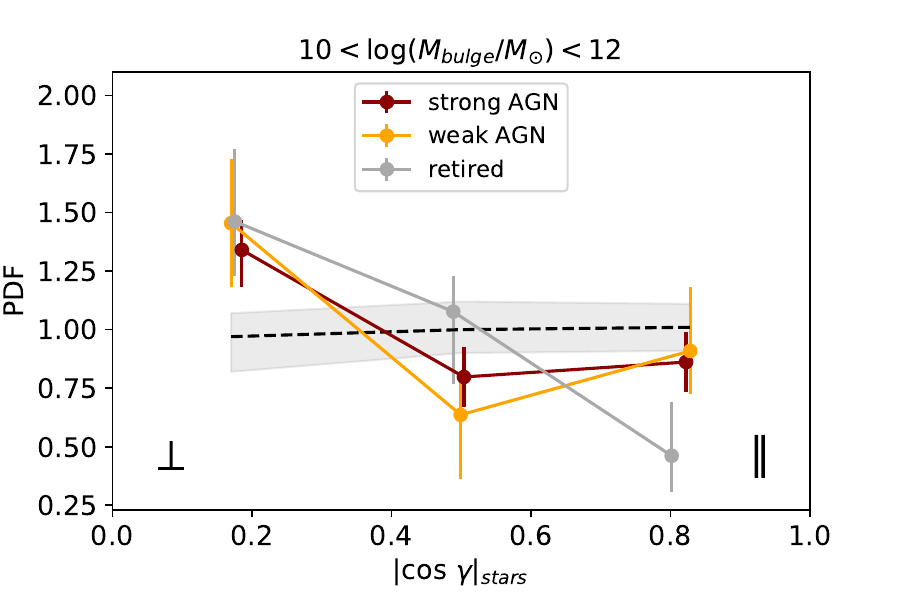}
\includegraphics[width=\columnwidth]{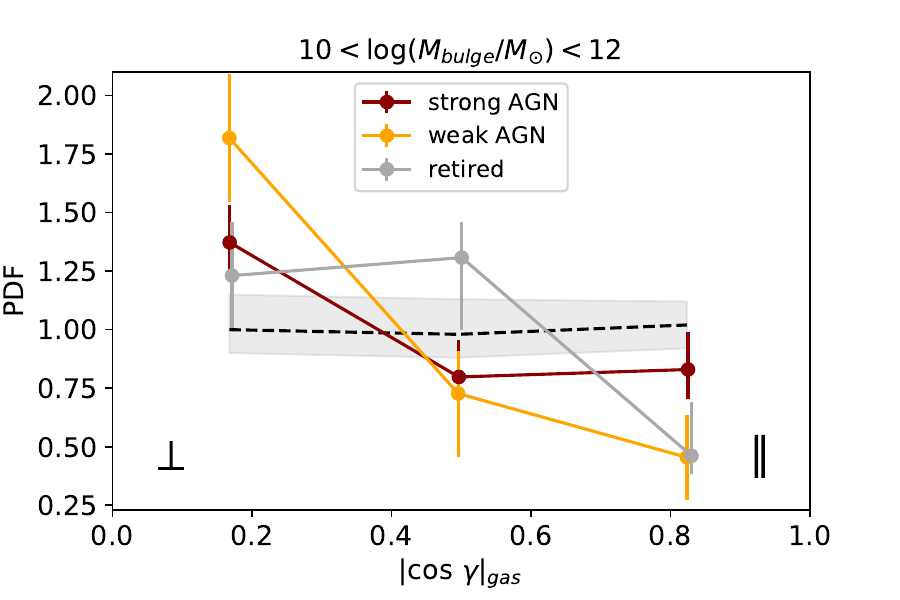}
\caption{PDFs of |$\cos\gamma$| for retired galaxies, weak AGN and strong AGN with $10<\log{(M_{\rm bulge}/M_{\odot})}<12$. For the left and right panels the galaxy spin axes are identified using spatially-resolved stellar and ionised gas kinematics, respectively. The black dashed line and the shaded region represent the reconstructed uniform distribution with its 95\% confidence intervals. For stellar spin--filament alignments a significant perpendicular tendency is found only for retired galaxies, while for gas spin--filament alignments strong and weak AGN tend to have more perpendicular tendencies.}
\label{WHAN_results}
\end{figure*}

In addition to spin--filament alignments measured using stellar kinematics, we explore the tendencies of gas spin--filament alignments for the same galaxy populations. In fact, gas accretion has been found to be tied to spin--filament alignments (e.g., \citealp{Barsanti2022}) and AGN are more able to affect ionised gas kinematics than stellar kinematics (e.g., \citealp{Oh2022}). Comparing |$\cos\gamma$|$_{\rm stars}$ with |$\cos\gamma$|$_{\rm gas}$, we find that most ($\sim$65\%) of the kinematically misaligned galaxies, i.e. those having $\Delta{\rm PA}$\,=\,$|{\rm PA_{stars}-PA_{gas}}|$\,>\,30\degree in line with previous studies (e.g., \citealp{Duckworth2019}), are retired galaxies. This is expected since the stars--gas misalignment is more prominent for early-type/passive galaxies than late-type/star-forming galaxies (e.g., \citealp{Bryant2019}). The finding suggests that different tendencies can be found for the gas spin--filament alignments relative to the stellar alignments of the retired galaxies. 

The panels of Figure~\ref{WHAN_results} display the PDFs of |$\cos\gamma$| for the galaxy populations based on the WHAN diagram and with $10<\log{(M_{\rm bulge}/M_{\odot})}<12$, where the galaxy spin axes are identified using spatially-resolved stellar (left) and ionised gas kinematics (right). The $M_{\rm bulge}$ distributions for these three galaxy populations with $10<\log{(M_{\rm bulge}/M_{\odot})}<12$ are not significantly different according to the two-sample K-S test. Table~\ref{KS_Results} lists the results from the K-S test. For stellar spin--filament alignments a significant perpendicular tendency is found only for retired galaxies ($p_{\rm K-S}$\,=\,0.009), while for gas spin--filament alignments both strong and weak AGN have more perpendicular tendencies ($p_{\rm K-S}$\,=\,0.006 and $p_{\rm K-S}$\,=\,0.005 respectively). Thus, for galaxies with massive bulges, the tendency to perpendicular spin--filaments alignments for stars is driven by retired galaxies, while for ionised gas we detect influence by AGN.

The analysis of retired galaxies, weak AGN and strong AGN with $8<\log{(M_{\rm bulge}/M_{\odot})}<10$ reveals uniform PDFs for the three galaxy populations, both for stellar and ionised gas spin--filament alignments. This highlights that AGN activity has a more dramatic impact on
galaxies with massive bulges.

\subsubsection{Seyfert galaxies versus LINERs}
\label{Seyfert galaxies versus LINERs}

\begin{figure*}
\includegraphics[width=\columnwidth]{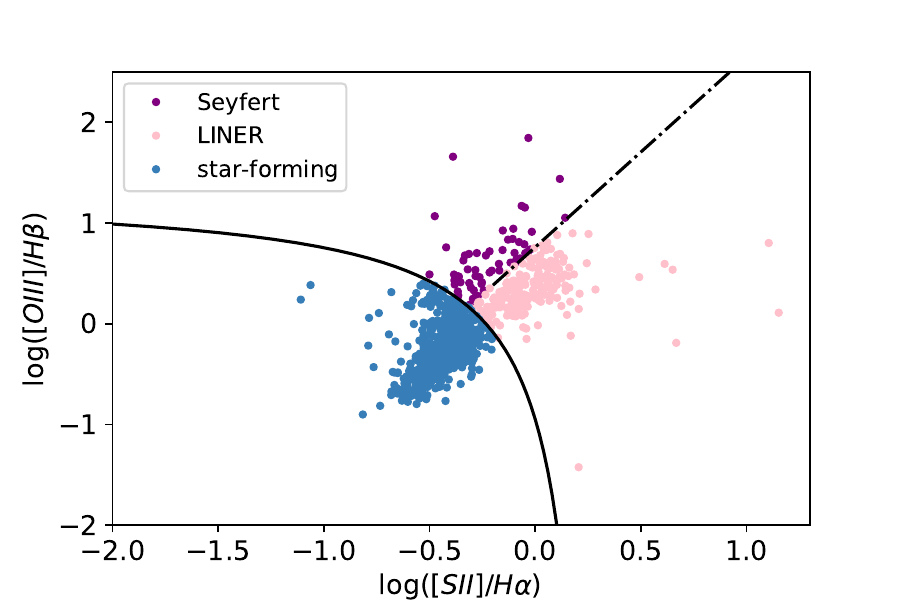}
\includegraphics[width=\columnwidth]{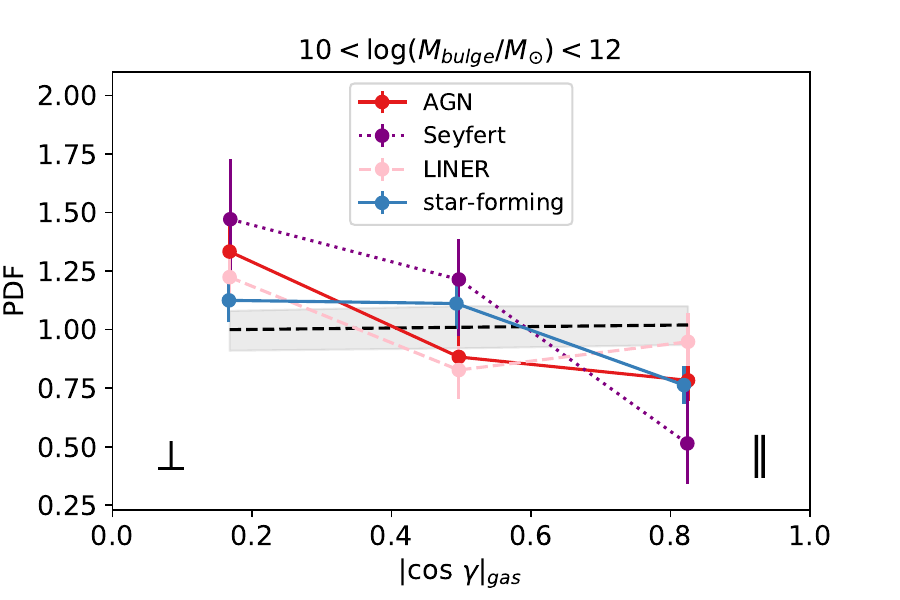}
\caption{Left: [S\,II]-based diagnostic diagram for Sample~A (977 SAMI galaxies). The solid line represents the theoretical maximal star-forming line of \citet{Kewley2001}. The dot-dashed line separates Seyfert galaxies from LINERs, as in \citet{Kewley2006}. Right: PDFs of gas spin--filament alignments for star-forming galaxies and AGN, divided into Seyferts and LINERs, with 10\,<\,$\log{(M_{\rm bulge}/M_{\odot})}$\,<\,12 and EW(H$\alpha$)\,>\,3\AA. The black dashed line and the shaded region represent the reconstructed
uniform distribution with its 95\% confidence intervals. The perpendicular tendency of AGN is dominated by Seyferts, while LINERs show a more uniform distribution. Seyferts also show a more perpendicular trend with respect to star-forming galaxies.}
\label{SII_results}
\end{figure*}

For AGN-like galaxies, the source of ionisation could be due to supermassive black holes (e.g., Seyfert galaxies) or to less hard/different mechanisms (e.g., shocks, evolved stellar populations). In empirical classifications, AGN-like galaxies might be identified as low-ionisation nuclear emission-line regions (LINERs; \citealp{Heckman1980}), having spectra dominated by emission lines from low-ionisation
species ([O I], [N II], [S II]). To investigate the influence of supermassive black holes on gas spin--filament alignments, here we classify galaxies from Sample~A using [S\,II]-based emission-line diagnostics, which are more effective in separating Seyfert galaxies from LINERs than [N\,II]-based diagnostics (such as the BPT and WHAN diagrams). The left panel of Figure~\ref{SII_results} displays log([O\,III]/H$\beta$) as a function of log([S\,II]/H$\alpha$), where the solid line is the theoretical maximal star-forming line of \citet{Kewley2001} and the dashed line separates Seyfert galaxies from LINERs according to \citet{Kewley2006}. We find that $\sim$30\% of Sample~A galaxies are not star-forming, with most being LINERs and only $\sim$20\% of AGN being Seyferts (as expected since the SAMI survey probes only the local Universe, where most AGN are LINERs).

To avoid the impact of $M_{\rm bulge}$, we select galaxies with 10\,<\,$\log{(M_{\rm bulge}/M_{\odot})}$\,<\,12 and to exclude retired galaxies we chose star-forming galaxies and AGN with EW(H$\alpha$)\,>\,3\AA. These criteria leave 224 star-forming galaxies and 190 AGN out of which $\sim$40\% are Seyfert galaxies. The right panel of Figure~\ref{SII_results} shows the PDFs of gas spin--filament alignments for these star-forming galaxies and AGN, divided into LINERs and Seyfert galaxies. The $M_{\rm bulge}$ distributions for LINERs and Seyferts with $10<\log{(M_{\rm bulge}/M_{\odot})}<12$ are not significantly different according to the two-sample K-S test. Table~\ref{KS_Results} displays the results from the K-S test. AGN have a more perpendicular tendency in agreement with the previous findings $p_{\rm K-S}=0.006$). Their signal is dominated by Seyfert galaxies ($p_{\rm K-S}=0.011$), while LINERs show a uniform distribution. Star-forming galaxies show a significant perpendicular tendency ($p_{\rm K-S}=0.020$) as expected for galaxies with massive bulges, but a weaker trend with respect to AGN and Seyferts. Similar findings are obtained for emission-line fluxes estimated within a more central aperture of 1.4\,arcsec. 

In conclusion, our results show that instantaneous AGN activity is linked to the gas spin--filament alignments.

\subsection{Role of central velocity dispersion in stellar versus gas spin--filament alignments}
\label{Role of central velocity dispersion in stellar versus gas spin-filament alignments}

Our results from Section~\ref{Impact of AGN} suggest that the instantaneous black hole activity might influence the ionised gas spin--filament alignments, while we do not detect any evidence in the stellar spin--filament alignments. In this Section, we explore the role of the integrated black hole activity, using central stellar velocity dispersion as proxy (see Section~\ref{Central velocity dispersions}). We investigate whether this galaxy parameter shows any correlation with the stellar and gas alignments, separately. 

\begin{figure}
\includegraphics[width=\columnwidth]{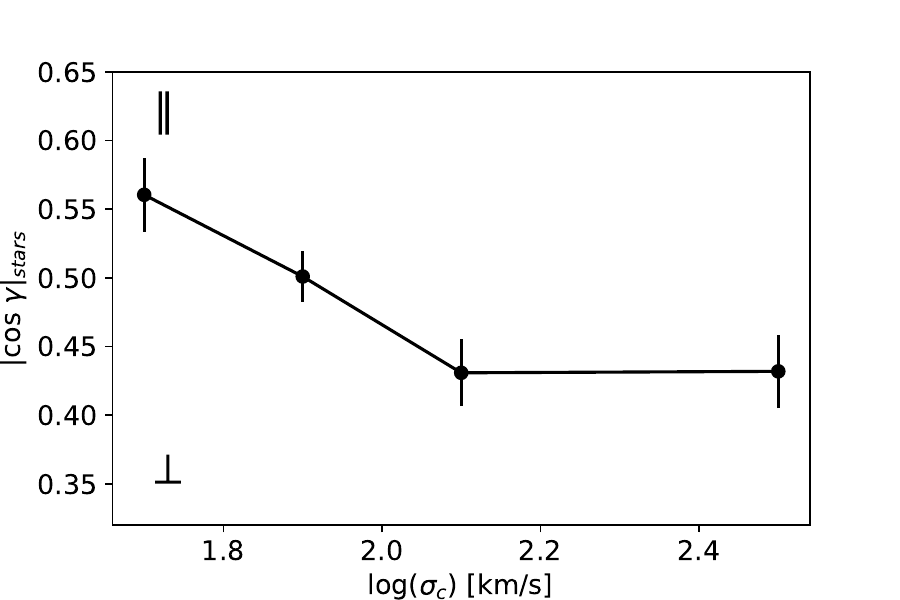}
\caption{Average $|\cos\gamma|_{\rm stars}$ values as a function of $\sigma_{\rm c}$ 
 for Sample~A (977 SAMI galaxies). The mean $|\cos\gamma|_{\rm stars}$ value decreases with increasing $\sigma_{\rm c}$, pointing to more perpendicular spin–filament alignments for higher $\sigma_{\rm c}$. A similar trend is found for $|\cos\gamma|_{\rm gas}$, but with lower statistical significance.}
\label{SigmaC_results}
\end{figure}

According to the Spearman test, for Sample~A we find that $|\cos\gamma|_{\rm stars}$ correlates with $\sigma_{\rm c}$ ($\rho$\,=\,$-$0.093 and $p_{\rm S}$\,=\,0.004). Figure~\ref{SigmaC_results} shows mean $|\cos\gamma|_{\rm stars}$ as a function of $\sigma_{\rm c}$: the average value of $|\cos\gamma|_{\rm stars}$ decreases with increasing $\sigma_{\rm c}$, pointing to more perpendicular spin–filament alignments for higher values of $\sigma_{\rm c}$. Consistent results are obtained for $\sigma_{\star}$. Similarly, $|\cos\gamma|_{\rm gas}$ correlates with $\sigma_{\rm c}$ ($\rho$\,=\,$-$0.069 and $p_{\rm S}$\,=\,0.037), but with lower statistical significance compared to $|\cos\gamma|_{\rm stars}$.  

A tight correlation between $\sigma_{\rm c}$ and $M_{\rm bulge}$ is found ($\rho$\,=\,0.834 and $p_{\rm S}$\,=\,$10^{-253}$). $M_{\rm bulge}$ shows a stronger correlation with $|\cos\gamma|_{\rm stars}$ ($\rho$\,=\,$-$0.127 and $p_{\rm S}$\,=\,$10^{-4}$) in comparison to $\sigma_{\rm c}$ for Sample~A. This suggests that $\sigma_{\rm c}$ might be a secondary tracer of spin--filament alignments with respect to $M_{\rm bulge}$. In the next section we investigate central velocity dispersion and bulge mass in more detail, estimating partial correlation coefficients and extending the analysis from spin--filament alignments to secular star formation quenching, to better understand the roles of the physical mechanisms traced by these parameters in galaxy evolution.

\subsection{Role of bulge mass and central velocity dispersion in spin--filament alignments versus quenching}
\label{Role of bulge mass and central velocity dispersion in spin--filament alignments versus quenching}

Velocity dispersion is found to be a key galaxy property correlating with quenching processes, especially for central galaxies and high-mass satellite galaxies \citep{Bluck2020,Brownson2022,Bluck2022}. In Section~\ref{Role of central velocity dispersion in stellar versus gas spin-filament alignments} we showed that central velocity dispersion is a secondary tracer of spin--filament alignments with respect to bulge mass. To shed light on the roles of these parameters in relation to the processes shaping galaxy evolution, we now explore correlations of central velocity dispersion and bulge mass for galaxy spin--filament alignments and star formation quenching. We perform analyses for Sample~B, i.e.\ those 845 SAMI galaxies from Sample~A that have measurements of star formation rates (see Section~\ref{Galaxy samples}). Sample B includes group central galaxies, group satellite galaxies, and isolated galaxies. We report the results for SED-fitting SFRs, but consistent findings are obtained for the spectroscopic SFR estimates.

To assess possible dependencies, we perform the Spearman rank correlation test for individual galaxies. In addition to bulge mass and central velocity dispersion, we also report the results in Table~\ref{SpearmanResults} for bulge-to-total flux ratio and stellar mass. These latter parameters have also been suggested as playing major roles in both galaxy spin--filament alignments \citep{Dubois2014,Laigle2015,Codis2018,Wang2018,Kraljic2020,Welker2020,Barsanti2022} and star formation quenching \citep{Baldry2006,Peng2010,Bluck2019}. As expected, statistically significant correlations with |$\cos\gamma$|$_{\rm stars}$, $|\cos\gamma|_{\rm gas}$ and $\Delta$(SFR) are detected for all parameters. The highest-significance correlation for |$\cos\gamma$|$_{\rm stars}$ and $|\cos\gamma|_{\rm gas}$ is found to be with $M_{\rm bulge}$, while for $\Delta$(SFR) it is with $\sigma_{\rm c}$.

\begin{figure*}
\includegraphics[scale=0.26]{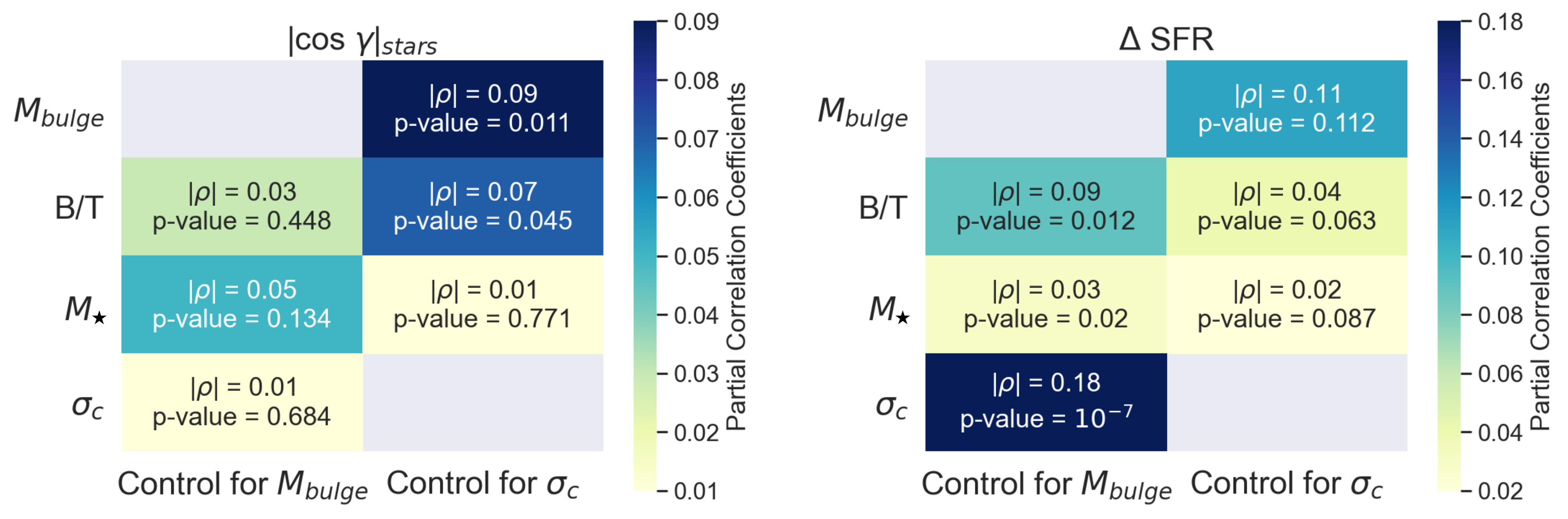}
\caption{Partial correlation coefficients from the Spearman test for |$\cos\gamma$|$_{\rm stars}$ (left panel) and $\Delta$(SFR) (right panel) of Sample~B (845 SAMI galaxies); in each case, we control for $M_{\rm bulge}$ (left column) and $\sigma_{\rm c}$ (right column). Darker colours indicate stronger residual correlations. For spin--filament alignments no correlation residuals with the other properties are found when controlling for $M_{\rm bulge}$, while for star formation quenching no significant residuals are detected when controlling for $\sigma_{\rm c}$.}
\label{PartialCorrelations_SFR_Angle}
\end{figure*}

To understand whether $M_{\rm bulge}$ and $\sigma_{\rm c}$ represent the primary parameter of correlation for |$\cos\gamma$|$_{\rm stars}$ and $\Delta$(SFR) respectively, we estimate the partial correlation coefficients (see Section~\ref{Correlation and causality tests}). For this and the following analyses, we report the findings only for |$\cos\gamma$|$_{\rm stars}$, since consistent results are found for $|\cos\gamma|_{\rm gas}$. The left panel of Figure~\ref{PartialCorrelations_SFR_Angle} shows the partial correlation coefficients from the Spearman test between |$\cos\gamma$|$_{\rm stars}$ and each of the studied galaxy properties while controlling for $M_{\rm bulge}$ or for $\sigma_{\rm c}$; the right panel shows the same for $\Delta$(SFR). For |$\cos\gamma$|$_{\rm stars}$ no significant correlations remain once the correlation with $M_{\rm bulge}$ is taken into account, whereas a significant residual is measured for $M_{\rm bulge}$ while controlling for $\sigma_{\rm c}$. In contrast, for $\Delta$(SFR) there is a significant dependency on $\sigma_{\rm c}$ while controlling for $M_{\rm bulge}$, whereas no residuals remain while controlling for $\sigma_{\rm c}$.

\begin{table}
\centering
\caption{Results from the Spearman rank correlation test and PLS regression for stellar spin--filament alignments (top table), ionised gas spin--filament alignments (middle table), and star formation quenching (bottom table) for Sample~B (845 galaxies). Column~1 gives the galaxy property analysed, column~2 the Spearman correlation coefficient, column~3 the $p$-value from the Spearman test (with $p$-values less than 0.05 highlighted in bold), and column~4 the explained variance in $|\cos\gamma|_{\rm stars}$, $|\cos\gamma|_{\rm gas}$ or $\Delta$(SFR).}
\label{SpearmanResults}
\begin{tabular}{@{}lccS@{}}
\multicolumn{4}{c}{$|\cos\gamma|_{\rm stars}$} \\
\toprule
Galaxy property                 & $\rho$  & $p_{\rm S}$        & {Variance (\%)} \\
\toprule
$\log(M_{\rm bulge}/M_{\odot})$ & $-$0.12 & $\mathbf{10^{-4}}$ & 78.8 \\
B/T                             & $-$0.11 & \textbf{0.001}     & 11.2 \\
$\log(M_\star/M_{\odot})$       & $-$0.08 & \textbf{0.018}     & 3.6 \\
$\sigma_{\rm c}$                    & $-$0.10 & \textbf{0.004}     & 6.4 \\
\midrule
\\
\multicolumn{4}{c}{$|\cos\gamma|_{\rm gas}$} \\
\midrule
Galaxy property                 & $\rho$  & $p_{\rm S}$        & {Variance (\%)} \\
\toprule
$\log(M_{\rm bulge}/M_{\odot})$ & $-$0.11 & $\mathbf{10^{-3}}$ & 75.6 \\
B/T                             & $-$0.10 & \textbf{0.002}     & 15.3 \\
$\log(M_\star/M_{\odot})$       & $-$0.07 & \textbf{0.015}     & 7.0 \\
$\sigma_{\rm c}$                    & $-$0.07 & \textbf{0.037}     & 2.0 \\
\midrule
\\
\multicolumn{4}{c}{$\Delta$(SFR)} \\
\toprule
Galaxy property                 & $\rho$  & $p_{\rm S}$         & {Variance (\%)} \\
\midrule
$\log(M_{\rm bulge}/M_{\odot})$ & $-$0.49 & $\mathbf{10^{-53}}$ & 11.0 \\
B/T                             & $-$0.49 & $\mathbf{10^{-52}}$ &  3.0 \\
$\log(M_\star/M_{\odot})$       & $-$0.34 & $\mathbf{10^{-23}}$ &  1.2 \\
$\sigma_{\rm c}$                    & $-$0.51 & $\mathbf{10^{-56}}$ & 84.8 \\
\bottomrule
\end{tabular}
\end{table}

\begin{figure*}
\includegraphics[scale=0.26]{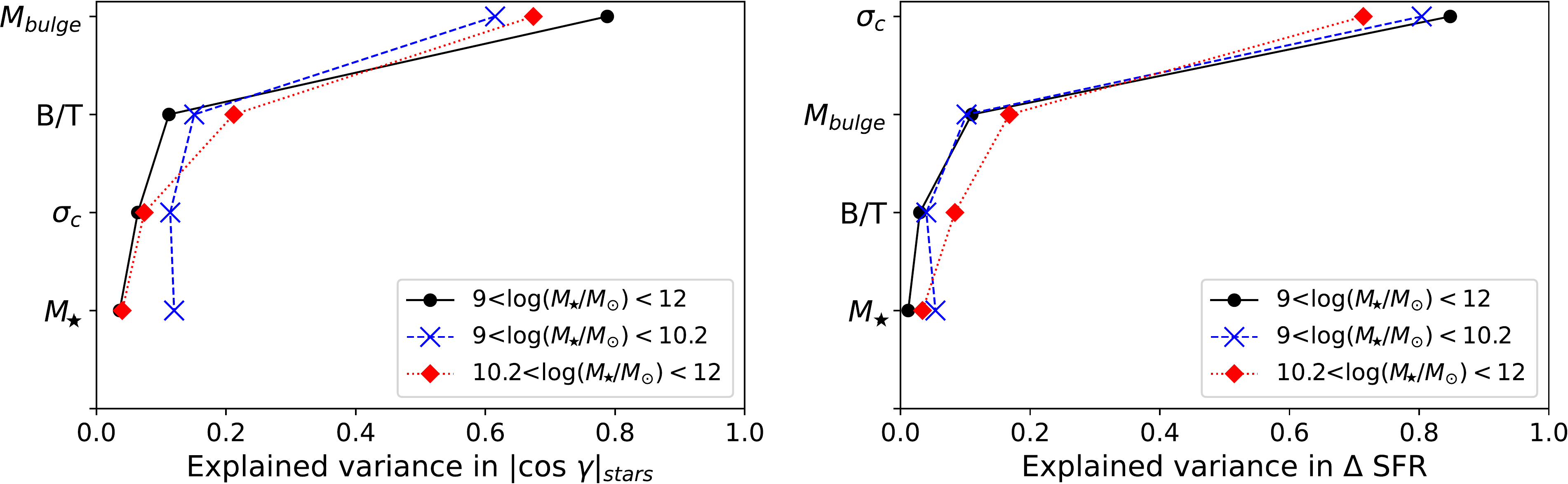}
\caption{Fraction of the variance in |$\cos\gamma$|$_{\rm stars}$ (left) and in $\Delta$(SFR) (right) explained by each galaxy property according to PLS regression applied to Sample~B (845 SAMI galaxies; black lines): $M_{\rm bulge}$ explains 79\% of the variance in |$\cos\gamma$|$_{\rm stars}$, while $\sigma_{\rm c}$ explains 85\% of the variance in $\Delta$(SFR). Similar trends are also found when dividing Sample~B into high-$M_{\star}$ and low-$M_{\star}$ subsamples.}
\label{PLS_SFR_Angle}
\end{figure*}

\begin{table*}
\centering
\caption{Results from Random Forest classifier for galaxy spin--filament alignments and star formation quenching. Column~1 lists the studied process, column~2 the total number of galaxies, column~3 the classification, column~4 the Area Under the Curve values for the training sample and column~5 the {\fontfamily{qcr}\selectfont min\_samples\_leaf} parameter.}
\label{RFresults}
\begin{tabular}{@{}lcccc@{}}
\toprule
Process & $N_{\rm gal}$ & Classification ($N_{\rm gal}$) & AUC$_{\rm train}$  & {\fontfamily{qcr}\selectfont min\_samples\_leaf}\\
\midrule
$|\cos\gamma|_{\rm stars}$ & 514 &  $\parallel$:\hspace{2mm}|$\cos\gamma$|$_{\rm stars}>0.7$ (278) & 0.78 & 30 \\
& & $\perp$:\hspace{2mm}|$\cos\gamma$|$_{\rm stars}<0.3$ (236)  \\
\\
$\Delta$(SFR) & 514 & PAS: $\Delta$(SFR) $<-1.0$ (298) &0.87 & 15 \\
&  & SF:\hspace{2mm} $\Delta$(SFR) $>-0.5$ (216)  \\
\bottomrule
\end{tabular}
\end{table*}

An alternative approach is to apply the partial least squares regression technique (PLS; see Section~\ref{Correlation and causality tests}). Figure~\ref{PLS_SFR_Angle} and Table~\ref{SpearmanResults} show the fraction of the variance in |$\cos\gamma$|$_{\rm stars}$ and $\Delta$(SFR) explained by each galaxy parameter. The highest contribution (79\%) to the variance in |$\cos\gamma$|$_{\rm stars}$ is found for $M_{\rm bulge}$, suggesting that this parameter most significantly correlates with galaxy spin--filament alignments. On the other hand, most (85\%) of the variance in $\Delta$(SFR) is explained by $\sigma_{\rm c}$, implying that this is the main parameter of correlation with star formation quenching. Dividing Sample~B into 345 low-mass galaxies with $9<\log{(M_\star/M_{\odot)}}<10.2$ and 500 high-mass galaxies with $10.2<\log{(M_\star/M_{\odot)}}<12$, we find similar trends in both ranges of stellar mass (see Figure~\ref{PLS_SFR_Angle}).


Finally, we apply the Random Forest classifier (see Section~\ref{Correlation and causality tests} for more details) to predict spin--filament alignments and star formation quenching separately for Sample~B. The algorithm is able to estimate the relative importance of each input galaxy parameter in predicting these processes. Specifically, the input galaxy parameters we investigate are bulge mass (as a merger tracer), central velocity dispersion (as a proxy for the black hole mass and its integrated activity), and stellar mass (related to supernova feedback as quenching mechanism and to the overall accretion within the cosmic web). We also assign a random number to each galaxy to keep track of the algorithm's performance. 

From Sample~B we identify the spin--filament alignment classification by selecting 278 galaxies with |$\cos\gamma$|$_{\rm stars}>0.7$ having a clear parallel alignment and 236 galaxies with |$\cos\gamma$|$_{\rm stars}<0.3$ having a clear perpendicular alignment. For the quenching classification, these 514 galaxies are divided into 298 passive galaxies with $\Delta$(SFR) $<-1.0$ and 216 star-forming galaxies with $\Delta$(SFR) $>-0.5$. This sample is largely ($\sim$60\%) central galaxies (i.e.\ the most massive galaxies within 0.25\,$R_{\rm 200}$ from the group centre as defined by \citealp{Santucci2020}) plus high-mass satellite galaxies (i.e.\ any other group member with $\log(M_\star/M_{\odot})>10$). 

\begin{figure}
\includegraphics[scale=0.27]{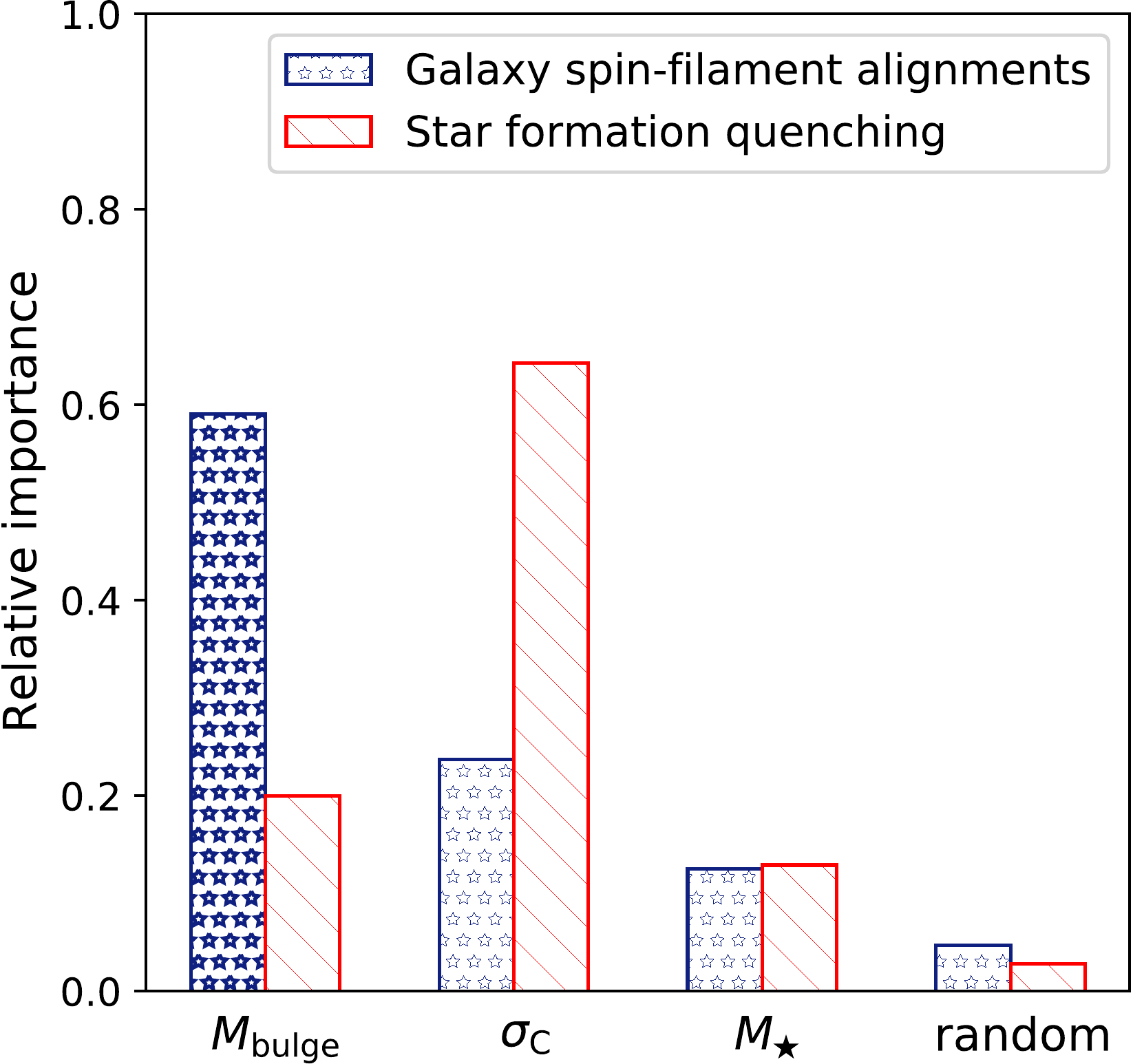}
\caption{Relative importance of each galaxy parameter for the galaxy spin--filament alignment classification (blue histogram) and the quenching classification (red histogram) from the Random Forest classifier. The classifications are based on 514 SAMI galaxies from Sample~B. $M_{\rm bulge}$ is the most important galaxy property to distinguish galaxies with more perpendicular alignments from those with more parallel alignments, while $\sigma_{\rm c}$ is the most important parameter for separating passive and star-forming galaxies.}
\label{RFfigure}
\end{figure}

For each classification, half the sample is used as the training sample while the other half is the test sample for which the algorithm estimates the respective predicted classification and quantifies its accuracy.  Examples of decision trees are shown in Appendix~\ref{Examples of decision trees from Random Forest analysis}. Table~\ref{RFresults} reports the Area Under the Curve (AUC) value for the training sample and the {\fontfamily{qcr}\selectfont min\_samples\_leaf} parameter chosen to maximise AUC$_{\rm train}$ with the constraint of AUC$_{\rm train}-$AUC$_{\rm test}<0.01$ to avoid over-fitting \citep{Piotrowska2022}. For both classifications the algorithm performs well with values of AUC$_{\rm train}>0.75$ (AUC values closer to 1 mark better classifiers). The Random Forest classifier measures the relative importance of each input galaxy parameter for determining the two classifications, as shown in Figure~\ref{RFfigure}. $M_{\rm bulge}$ and $\sigma_{\rm c}$ are found to be the most important galaxy properties determining the galaxy spin--filament alignment classification and the quenching classification, respectively. The random number scores the lowest importance for both processes (<5\%), highlighting the high accuracy of the classifiers. 

Overall, these tests demonstrate that $M_{\rm bulge}$ is the primary galaxy property correlating with spin--filament alignments, while $\sigma_{\rm c}$ is the best predictive parameter for star formation quenching. 

\section{Discussion}
\label{Discussion}

The aim of this paper is to shed light on the possible influence of black hole activity on the alignment of the galaxy spin axis with respect to the orientation of the closest cosmic filament. Here we address our results with respect to previous studies and their possible physical information. Finally, since machine learning is a rapidly evolving field, we discuss some related topics.

\subsection{Influence of instantaneous AGN activity}
\label{Influence of instantaneous black hole activity}

To our knowledge, this is the first study using observations to investigate the role of AGN in affecting galaxy spin--filament alignments. Our results from Section~\ref{Impact of AGN} confirm that $M_{\rm bulge}$, a tracer of mergers \citep{Sales2012,Wilamn2013}, is the primary parameter of correlation. This is in agreement with the picture where mergers are the main physical mechanisms causing spin--filament flips from parallel to perpendicular (e.g., \citealp{Codis2012,Dubois2014,Welker2014,Welker2020,Barsanti2022}). In addition to bulge growth and spin--filament alignment flips, mergers are also found to be able to trigger supermassive black holes via the strong gravitational
torque that fuels gas towards the inner region of galaxies \citep{Barnes1991,Hopkins2006,Hopkins2008,Gao2020}. Thus, mergers can be identified as the narrative thread among spin--filament alignments, bulge accretion and AGN activity.  

The study of stellar spin--filament alignments reveals that active galaxies are significantly more likely to have perpendicular alignments because they tend to be found in galaxies with higher bulge masses. This perpendicular tendency for galaxies with massive bulges is dominated by retired galaxies rather than weak or strong AGN, meaning that we do not see any impact from instantaneous AGN activity. These results are in agreement with the study of \citet{Soussana2020}, who found stronger perpendicular alignments in simulations when AGN feedback is implemented versus no AGN feedback, as a consequence of the larger percentage of massive pressure-supported galaxies that populate simulations with AGN-mode feedback.

On the other hand, ionised gas spin--filament alignments for AGN-like galaxies with high bulge masses tend to be more perpendicular than retired galaxies in the same bulge mass range. In particular, this tendency is seen for Seyfert galaxies compared to LINERs. For low-$M_{\rm bulge}$ galaxies, we find uniform PDFs for all these galaxy populations, highlighting that AGN activity has a more dramatic impact on galaxies with massive bulges, a finding that has also been observed by \citet{Soussana2020}. 
These results suggest that when the gas spin--filament alignment is parallel, gas accretion occurs mostly in the outer region of the galaxy, preferentially triggering star formation with little or no gas reaching the innermost region, thus there is no accretion for the supermassive black hole. In contrast, when the gas spin--filament alignment is perpendicular, this may favour gas falling and being torqued towards the centre, making AGN events more likely. As a consequence, AGN can contribute to maintaining a strong perpendicular tendency for the gas spin--filament alignments, preserving the flipped spin orientation after the merger. These findings support the scenario proposed by \citet{Dubois2014}, in which mergers trigger the flipping of the galaxy spin--filament alignment from parallel to perpendicular, and subsequent AGN activity acts to preserve the perpendicular tendency by preventing gas accretion and new gas inflow, which would lead to realignment back to a parallel orientation.

A link between AGN and spin--filament alignments is found only for the ionised gas. This is in agreement with the fact that AGN events have very fast duty cycles (1-10 Myr) and might impact gas kinematics (e.g., \citealp{Oh2022}). On the other hand, stellar kinematics takes Gyr-scales to change and such short-lived events might not be able to leave traces in it. Gas outflows driven by AGN feedback, which are able to stop in-situ star formation (e.g., \citealp{Dubois2013}) and thus accretion, are one possible mechanism responsible for the kinematic misalignment between the rotation axes of stars and gas \citep{Ristea2022}. In addition, \citet{Raimundo2023} observed that galaxies with kinematic misalignment show an increase in black hole activity, in agreement with our finding of a significant influence by AGN on gas alignments in comparison to stellar spins. Finally, \citet{Malavasi2021} found for the IllustrisTNG simulation that galaxy spin-related quantities show stronger trends with cosmic web distances when the spin is estimated only from the gas. In particular, they showed that the flipping of the spin--filament alignments from parallel to perpendicular as galaxies move along filaments is driven by the gas component. This is consistent with our results that tie spin--filament alignments to accretion and to mergers, since these latter, through the destruction and re-formation of gas discs, trigger bulge growth and the alignment flips. 

\subsection{AGN identification}
\label{AGN identification}

In this work we make use of emission-line diagnostic plots to identify AGN (Section~\ref{Diagnostics of AGN}). Classifications of AGN in different electromagnetic bands allow us to study the different physical mechanisms related to these complex objects (see \citealt{Padovani2017} for a review), but none of these are foolproof with regard to contamination of star-forming galaxies and shocks. Combining multi-wavelength classifications for the same galaxies might find only a small overlapping AGN sample (e.g, \citealp{Leahy2019}).

To explore X-ray classification of AGN and its impact on galaxy spin--filament alignments, we analyse whether galaxies with massive bulges and X-ray AGN tend to have more perpendicular alignments relative to those with no X-ray AGN detection, pointing to a possible role of the AGN activity in maintaining the perpendicular alignment. We take advantage of the X-ray AGN catalogue from the eROSITA Final Equatorial-Depth Survey \citep{Liu2021}, which covers the GAMA G09 region. 

\begin{figure}
\includegraphics[width=\columnwidth]{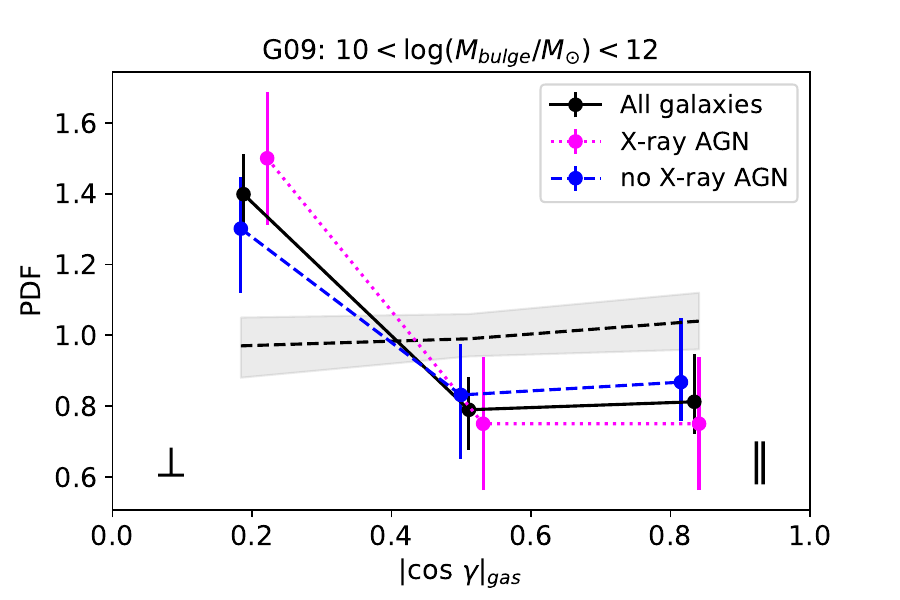}
\caption{PDFs of gas spin--filament alignments for 131 galaxies from G09 with $10<\log{(M_{\rm bulge}/M_{\odot})}<12$, divided into 48 galaxies with X-ray AGN and 83 with no X-ray AGN detection. The black dashed line and the shaded region represent the reconstructed
uniform distribution with its 95\% confidence intervals. The perpendicular alignments for the galaxies with massive bulges are mainly due to X-ray AGN.}
\label{Xray_results}
\end{figure}

From Sample~A, we select 131 galaxies in G09 with $10<\log{(M_{\rm bulge}/M_{\odot})}<12$. Of these, 48 galaxies are classified as X-ray AGN, while the remaining 83 are not X-ray AGN candidates. The $M_{\rm bulge}$ distributions for these two populations are not significantly different according to the two-sample K-S test. Figure~\ref{Xray_results} displays the PDFs of gas spin--filament alignments, showing that galaxies with massive bulges and X-ray AGN have the strongest perpendicular alignment with respect to the reconstructed uniform distribution ($p_{\rm K-S}=0.012$; see the K-S test results in Table~\ref{KS_Results}). Similar trends are observed for the stellar spin--filament alignments, but with weaker statistical significance. These findings support the results of Sections~\ref{AGN influence gas} and~\ref{Seyfert galaxies versus LINERs} based on emission-line diagnostic AGN, suggesting that the AGN activity acts to maintain the perpendicular spin--filament alignments of galaxies with massive bulges.

We also investigate other AGN classifications exploiting the radio LARGESS data \citep{Ching2017} and the infrared WISE data \citep{Yao2020}, but we find only 10 and 16 AGN respectively, insufficient to obtain statistically significant results. Larger spectroscopic galaxy samples, such as the Wide Area VISTA Extra-Galactic Survey (WAVES; \citealp{Driver2016}) at $z<1$ and the Main MOONS GTO Extragalactic Survey (MOONRISE; \citealp{Maiolino2020}) at $1<z<2$, together with multi-wavelength studies of AGN, are needed to explore the role of AGN activity in shape--filament alignments based on photometric position angles. 

\subsection{\hmath{\sigma_{\rm c}}--\hmath{M_{\rm bulge}} plane: integrated AGN feedback and mergers}
\label{Impact of integrated AGN feedback}

Investigating the dominant physical mechanisms involved in specific processes of galaxy evolution by exploring the strongest correlations with galaxy parameters has become a standard tool of research in observational astrophysics. Different galaxy properties trace different physical mechanisms and may be associated with different processes of galaxy evolution. In this work we study how bulge mass (a tracer of mergers) and central velocity dispersion (a tracer of integrated black hole activity) relate to galaxy spin--filament alignments and star formation quenching.

 \begin{figure*}
 \includegraphics[scale=0.285]{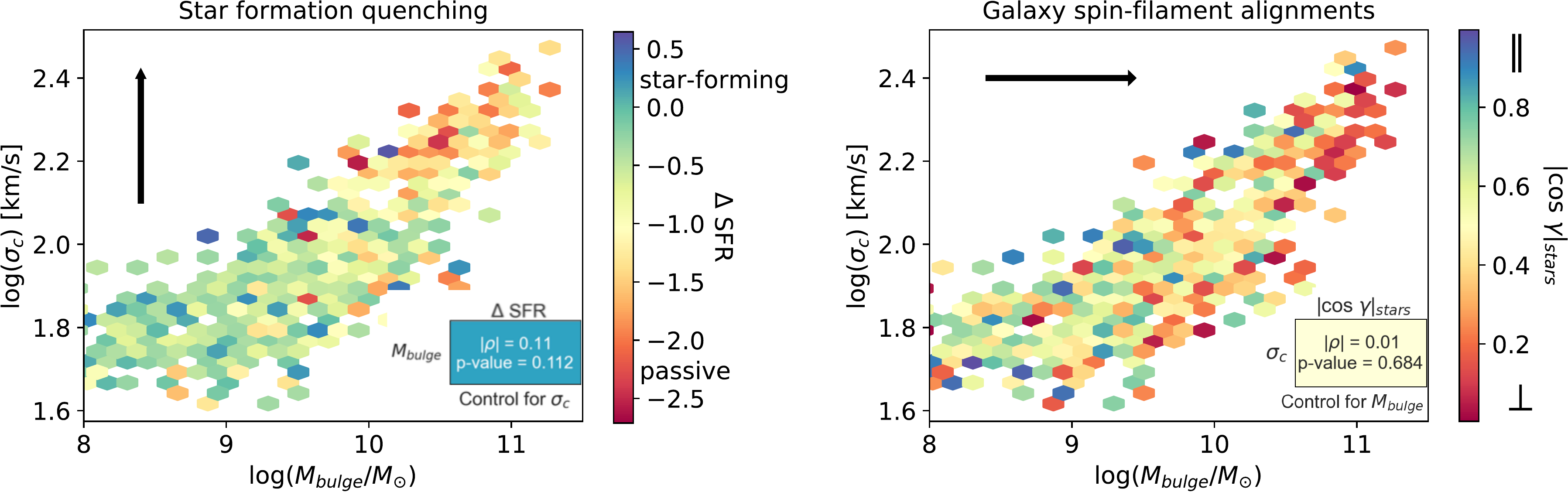}
\caption{$\sigma_{\rm c}$--$M_{\rm bulge}$ plane for Sample~B (845 SAMI galaxies) divided into hexagonal bins and colour-coded according to the mean values of the logarithmic offsets from the stellar mass–star formation rate main sequence $\Delta$(SFR) (left) and stellar spin--filament alignments |cos $\gamma$|$_{\rm stars}$ (right). The mean spin--filament alignment appears to vary horizontally (i.e.\ arrow following $M_{\rm bulge}$), while star formation quenching appears to vary vertically (i.e.\ arrow following $\sigma_{\rm c}$). These visual impressions are in qualitative agreement with the quantitative results from the Random Forest classifier in Section~\ref{Role of bulge mass and central velocity dispersion in spin--filament alignments versus quenching}. The inserts show no significant partial correlation coefficients for $\Delta$(SFR) versus $M_{\rm bulge}$ controlling for $\sigma_{\rm c}$ (left) and |cos $\gamma$|$_{\rm stars}$ versus $\sigma_{\rm c}$ controlling for $M_{\rm bulge}$ (right), taken from Figure~\ref{PartialCorrelations_SFR_Angle}.}
\label{Sigma_Mbulge_plane}
\end{figure*}

Galaxy stellar mass is found to play a key role in both star formation quenching \citep{Baldry2006,Peng2010,Bluck2019} and galaxy spin--filament alignments \citep{Dubois2014,Laigle2015,Codis2018,Wang2018,Kraljic2020,Welker2020}. Bulge stellar mass is observed to show an even stronger correlation with both star formation quenching \citep{Lang2014,Bluck2014,Bluck2022,Dimauro2022} and galaxy spin--filament alignments \citep{Barsanti2022}. However, \citet{Bluck2022} found that for star formation quenching the correlation with bulge mass disappears once the correlation with central velocity dispersion is taken into account, while \citet{Barsanti2022} found that bulge mass is the primary parameter of correlation with galaxy spin--filament alignments, with no residual left for other galaxy properties to explain. In this work, we have verified that the latter result also stands for central velocity dispersion (Section~\ref{Role of central velocity dispersion in stellar versus gas spin-filament alignments}). Using Random Forest classification on a consistent galaxy sample, we confirm that bulge mass is the most predictive property for galaxy spin--filament alignments, while central velocity dispersion is the primary predictor for star formation quenching (Section~\ref{Role of bulge mass and central velocity dispersion in spin--filament alignments versus quenching}).

Following \citet{Bluck2022}, we provide a visual confirmation of our results in Figure~\ref{Sigma_Mbulge_plane}. Both panels display the $\sigma_{\rm c}$--$M_{\rm bulge}$ plane for Sample~B (845 SAMI galaxies), divided into hexagonal bins and colour-coded according to the mean values of the logarithmic offsets from the stellar mass–star formation rate main sequence $\Delta$(SFR) (left) and the stellar spin--filament alignments |cos $\gamma$|$_{\rm stars}$ (right). Inspection by eye suggests that star formation quenching varies vertically in the plane, i.e.\ following $\sigma_{\rm c}$ and in agreement with Figure~6 of \citet{Bluck2022}. In contrast, the flipping of spin--filament alignments proceeds horizontally, i.e.\ following $M_{\rm bulge}$. The right panel is noisier than the left panel, underlining that spin--filament alignments are a weak signal at low redshifts ($z<0.1$; \citealp{Codis2018}), representing a memory of the galaxy's formation. As expected, since more bulge-dominated galaxies are also characterised by higher dispersion, $\sigma_{\rm c}$ and $M_{\rm bulge}$ are tightly correlated ($\rho=0.822$ and $p_{\rm S}=10^{-157}$, in agreement with Section~\ref{Role of central velocity dispersion in stellar versus gas spin-filament alignments} and with Section~4.2.2 of \citealt{Bluck2022}). The average root-mean-square dispersion is 0.10 dex, implying a strong physical connection between the two galaxy properties. However, the fact that $\sigma_{\rm c}$ and $M_{\rm bulge}$ are separately found to be the primary correlation parameters with two different processes in galaxy evolution suggests that this scatter encodes physical information. 

We extend the analysis of the roles of bulge mass and central velocity dispersion in star formation quenching to the whole GAMA equatorial regions within $z<0.1$ (i.e.\ including GAMA galaxies without SAMI data) in Appendix~\ref{Role of bulge mass and central velocity dispersion in quenching for GAMA galaxies}. We use the central velocity dispersion measurements from the 1D GAMA spectra ($\sigma_{\star}$; see Section~\ref{Central velocity dispersions}). Applying the Random Forest classifier, we find that $\sigma_{\star}$ is the most predictive parameter for quenching, in agreement with the results from Section~\ref{Role of bulge mass and central velocity dispersion in spin--filament alignments versus quenching} and previous studies \citep{Bluck2020,Bluck2022,Piotrowska2022}. These latter works observed that it is the star formation quenching of centrals and high-mass satellite galaxies that is predominantly predicted by central velocity dispersion. Our GAMA galaxy sample is mainly (66\%) comprised of centrals and high-mass satellites with $\log(M_\star/M_{\odot})>10$. Selecting only these galaxy populations, the relative importance of $\sigma_{\star}$ increases from 0.58 to 0.7 (see Figure~\ref{RFfigureGAMA}), in line with the relative importance of 0.8 for central velocity dispersion and black hole mass found by \citet{Bluck2022} and \citet{Piotrowska2022}, respectively; the slightly higher relative importance may be due to the much larger galaxy samples employed in those studies. 

Overall, these findings suggest that the dominant physical mechanisms that cause star formation quenching and flipping of the galaxy spin--filament alignments are different: AGN feedback drives galaxy quenching, while mergers trigger spin--filament alignment flips and bulge growth. These results link bulge mass and central velocity dispersion to separate processes of galaxy evolution, pointing to a possible physical explanation for the scatter in the $\sigma_{\rm c}$--$M_{\rm bulge}$ plane.

\subsection{Gradient boosted decision trees, model explainability and collinearity}

In this work we make use of the Random Forest algorithm. Another ensemble machine learning technique that combines multiple decision trees to create a robust predictive model is gradient boosting, where decision trees are iteratively added to correct the errors of the previous trees, gradually improving the model's performance. Like Random Forest, Gradient Boosted Decision Trees can be used for both regression and classification tasks. One gradient boosting implementation is the XGBoost package (eXtreme Gradient Boosting; \citealp{ChenG16}). XGBoost models tend to outperform Random Forest but are more likely to overfit the dataset they were trained on, particularly for small training sets like those studied in this work.

Model explainability or interpretability is a critical aspect of machine learning, meaning the ability to understand and interpret how a model arrives at its decisions or predictions. In this work, we use Gini impurity to assess the importance of the features, though we obtain the same outcomes using the permutation technique (see Section~\ref{Correlation and causality tests}). Both methods have some downsides, and in particular do not consider relationships between features. Recently, to overcome these limitations, \citet{Lundberg2020} introduce the SHAP (SHapley Additive exPlanations) package, which is an interpretability tool based on cooperative game theory and Shapley values. The main idea is to consider all possible combinations of features and calculate how much each feature contributes to the prediction or classification when combined with other features. SHAP is able to assess the interactions between  features and to provide both local (i.e. for a single measurement of the parameter of interest) and global (i.e. for the whole dataset) importances.

The correlations between features are referred as collinearity. Collinearity impacts the stability and feature importance of machine learning models. Random Forest is generally less affected by collinearity compared to logistic regression models that rely on linear relationships between features and the parameter of interest. Appendix B2 of \citet{Bluck2022} shows how the Random Forest Classifier is not affected by collinearity when using the All Parameter mode of operation, shifting from correlation to causation. However, for cases with low predictive power or highly correlated features, the stochastic nature of forest building means that two correlated features can be chosen one or the other when minimising the Gini impurity, leading to variable results. Given these caveats, we emphasise that care is always required when using machine learning tools.


\section{Summary and conclusions}
\label{Summary and conclusions}

\begin{table*}
\centering
\caption{Conceptual summary of the studied galaxy evolution processes, galaxy parameters and physical mechanisms.
Column~1 lists the galaxy evolution process, column~2 the primary galaxy parameter, column~3 the driving physical mechanism and columns~4 specifies the impact of black hole activity.}
\label{Conceptual_Table}
\begin{tabular}{@{}lccl@{}}
\toprule
Galaxy evolution process & Primary galaxy parameter & Driving physical mechanism & Impact of black hole activity \\
\midrule
Stellar spin--filament alignments &$M_{\rm bulge}$ & Mergers & No impact for instantaneous activity\\
& & & Secondary impact for integrated activity\\
\\
Ionised gas spin--filament alignments & $M_{\rm bulge}$& Mergers & Secondary impact for instantaneous activity\\
& & & Secondary impact for integrated activity\\
\\
 Star formation quenching &
$\sigma_{\rm c}$ & Integrated AGN feedback & Integrated activity is the main driver \\
\bottomrule
\end{tabular}
\end{table*}

In this study we use the SAMI Galaxy Survey to explore instantaneous and integrated black hole activity as factors in both stellar spin--filament alignments (|$\cos\gamma$|$_{\rm stars}$) and ionised gas spin--filament alignments (|$\cos\gamma$|$_{\rm gas}$). We analyse AGN-like galaxies identified from emission-line diagnostics, which offer a snapshot of the current activity state of the galaxies, allowing us to investigate the role of the instantaneous black hole activity. Central stellar velocity dispersion ($\sigma_{\rm c}$) is used as a proxy for the black hole mass and its integrated AGN feedback. We investigate the roles of $\sigma_{\rm c}$ and bulge mass ($M_{\rm bulge}$) in relation to spin--filament alignments and star formation quenching. Table~\ref{Conceptual_Table} summarises the studied galaxy evolution processes, galaxy parameters and physical mechanisms. Our main results are as follows:
\begin{enumerate}
    \item[(i)] C$_{\rm AGN}$, a proxy for the instantaneous AGN activity defined from the BPT diagram, correlates with |$\cos\gamma$|$_{\rm stars}$ ($\rho=-0.090$, $p_{\rm S}=0.005$) but shows a strong dependency on $M_{\rm bulge}$ ($\rho=-0.783$, $p_{\rm S}=10^{-202}$), and so is a secondary tracer of spin--filament alignments. Similar results are found for |$\cos\gamma$|$_{\rm gas}$, although with a higher contribution to the variance from C$_{\rm AGN}$ for ionised gas spin--filament alignments relative to |$\cos\gamma$|$_{\rm stars}$ (10\% versus 0.6\%). 
    \item[(ii)] The tendency to perpendicular values of |$\cos\gamma$|$_{\rm stars}$ for galaxies with massive bulges ($10<\log{(M_{\rm bulge}/M_{\odot})}<12$) is dominated by retired galaxies ($p_{\rm K-S}=0.009$), i.e.\ galaxies that have stopped star formation and whose ionisation is caused by evolved low-mass stars. On the other hand, for the same bulge mass range, |$\cos\gamma$|$_{\rm gas}$ shows a more perpendicular tendency for AGN-like galaxies ($p_{\rm K-S}=0.006$ and $p_{\rm K-S}=0.005$ for strong and weak AGN, respectively) than for retired galaxies. 
    \item[(iii)] For galaxies with $10<\log{(M_{\rm bulge}/M_{\odot})}<12$, Seyferts tend to have a more perpendicular |$\cos\gamma$|$_{\rm gas}$ alignment ($p_{\rm K-S}=0.011$) than LINERs, which show a more uniform distribution of alignments. 
    \item[(iv)] $\sigma_{\rm c}$, a proxy for integrated AGN feedback, shows a stronger correlation with |$\cos\gamma$|$_{\rm stars}$ ($\rho=-0.093$, $p_{\rm S}=0.004$) than |$\cos\gamma$|$_{\rm gas}$ ($\rho=-0.069$, $p_{\rm S}=0.037$). $M_{\rm bulge}$ shows a stronger correlation with $|\cos\gamma|_{\rm stars}$ ($\rho$\,=\,$-$0.127 and $p_{\rm S}$\,=\,$10^{-4}$) in comparison to $\sigma_{\rm c}$, pointing to $\sigma_{\rm c}$ to be a secondary tracer of spin--filament alignments. 
    \item[(v)] $M_{\rm bulge}$ is the primary parameter of correlation for galaxy spin--filament alignments, with no residual left for $\sigma_{\rm c}$. In contrast, $\sigma_{\rm c}$ is the most predictive parameter for star formation quenching, with no residual left for $M_{\rm bulge}$. These results are confirmed by estimating partial correlation coefficients, by using PLS regressions, and by implementing Random Forest classifications. 
\end{enumerate}

In conclusion, $M_{\rm bulge}$ is confirmed to be the primary tracer of galaxy spin--filament alignment flips, suggesting that mergers might be the main driver of the alignments. Mergers are also more likely to trigger AGN (e.g., \citealp{Gao2020}). Hence galaxies with perpendicular alignments can be characterised by a strong AGN activity which maintains the perpendicular tendency via preventing subsequent smooth gas accretion and re-alignment in parallel. This latter finding is observed for ionised gas spin--filament alignments, which show a link with the instantaneous AGN activity. For stellar spin--filament alignments we do not see any influence from instantaneous AGN activity, but a significant correlation is found for the integrated black hole activity. This latter is likely a secondary mechanism with respect to mergers. In fact, bulge mass and central velocity dispersion are each found to be the most predictive parameter for a different galaxy evolution process, with mergers triggering alignment flips and integrated AGN feedback driving star formation quenching. This highlights how bulge mass and central velocity dispersion, which have a correlation characterised by a meaningful physical scatter, play distinct roles in galaxy evolution.

This study provides new observational insights regarding the impact of black hole activity on galaxy spin--filament alignments. However, it is limited by a relatively small galaxy sample and consequently by the weak statistical significance of the results. Using SDSS, \citet{Lee2023} found a stronger statistical significance for the flipping of shape alignments of spiral galaxies with respect to the directions toward the void centers instead of the nearby filaments. In terms of the cosmic web reconstruction, \citet{Zakharova2023} found in the Millennium Simulation \citep{Springel2005} that filaments based on dark matter particles and galaxies show some deviations as a function of the local galaxy environments. 

Future IFS galaxy surveys, such as the Hector survey \citep{Bryant2020}, which provides spatially-resolved kinematics for stars and ionised gas for more than 10,000 galaxies in different environments, are needed to confirm the hints of the galaxy evolution mechanisms found in this work and to clarify the roles of cosmic structures and local environments in determining galaxy spins. Such IFS surveys in combination with multi-wavelength studies of AGN will deepen our understanding of the role of supermassive black holes and their activity timescale in influencing galaxy spin--filament alignments.

\section*{Acknowledgements}

We thank the referee for a constructive and helpful report. SB would like to thank Ángel R. López-Sánchez for insightful comments. This research was supported by the Australian Research Council Centre of Excellence for All Sky Astrophysics in 3 Dimensions (ASTRO 3D), through project number CE170100013. The SAMI Galaxy Survey is based on observations made at the Anglo-Australian Telescope. The Sydney-AAO Multi-object Integral field spectrograph (SAMI) was developed jointly by the University of Sydney and the Australian Astronomical Observatory, and funded by ARC grants FF0776384 (Bland-Hawthorn) and LE130100198. The SAMI input catalogue is based on data taken from the Sloan Digital Sky Survey, the GAMA Survey and the VST/ATLAS Survey. The SAMI Galaxy Survey is supported by the Australian Research Council Centre of Excellence for All Sky Astrophysics in 3 Dimensions (ASTRO 3D), through project number CE170100013, the Australian Research Council Centre of Excellence for All-sky Astrophysics (CAASTRO), through project number CE110001020, and other participating institutions. The SAMI Galaxy Survey website is http://sami-survey.org/. This study uses data provided by AAO Data Central (http://datacentral.org.au/). MMC acknowledges the support of a Royal Society Wolfson Visiting Fellowship (RSWVF\textbackslash R3\textbackslash 223005) at the University of Oxford. FDE acknowledges funding through the H2020 ERC Consolidator Grant 683184, the ERC Advanced grant 695671 ``QUENCH'' and support by the Science and Technology Facilities Council (STFC). SO acknowledges support from the NRF grant funded by the Korea government (MSIT) (No. 2020R1A2C3003769 and No. RS-2023-00214057).
JJB acknowledges support of an Australian Research Council Future Fellowship (FT180100231). AR acknowledges the receipt of a Scholarship for International Research Fees (SIRF) and an International Living Allowance Scholarship (Ad Hoc Postgraduate Scholarship) at The University of Western Australia. CW's contribution was supported in part by the National Science Foundation under Grant No. NSF PHY-1748958. JvdS acknowledges support from an Australian Research Council Discovery Early Career Research Award (DE200100461) funded by the Australian Government.

For this work we used the \href{http://www.python.org}{Python} programming language \citep{vanrossum1995}. We acknowledge the use of {\sc \href{https://pypi.org/project/numpy/}{numpy}} \citep{harris+2020}, {\sc \href{https://pypi.org/project/scipy/}{scipy}} \citep{jones+2001}, {\sc \href{https://pypi.org/project/matplotlib/}{matplotlib}} \citep{hunter2007}, {\sc \href{https://pypi.org/project/astropy/}{astropy}} \citep{astropyco+2013}, {\sc \href{https://pingouin-stats.org/}{pingouin}} \citep{Vallat2018}, {\sc \href{https://scikit-learn.org/stable/}{scikit-learn}} \citep{Pedregosa2012} and {\sc \href{http://www.star.bris.ac.uk/~mbt/topcat/}{topcat}} \citep{taylor2005}.

\section*{Data availability}

The SAMI reduced data underlying this article are publicly available at \href{https://docs.datacentral.org.au/sami}{SAMI Data Release 3} \citep{Croom2021}. Ancillary data comes from the \href{http://gama-survey.org}{GAMA Data Releases 3 and 4} \citep{Baldry2018,Driver2022}. The X-ray AGN catalogue from the eROSITA Final Equatorial-Depth Survey is available at \href{https://erosita.mpe.mpg.de/edr/eROSITAObservations/Catalogues/}{eROSITA-DE: Early Data Release site} \citep{Liu2021}. 

\section*{Author Contribution Statement}

SB devised the project, carried out the analysis and drafted the paper. FDE performed MGE fits and velocity dispersion estimates from 1D spectra for GAMA. SC performed 2D photometric bulge-disc decomposition. AR assembled the SED-fitting SFR catalogue. SB, MC, FDE, SO, YM, CW and HRMZ contributed to data analyses and interpretation of the results. JJB, SMC and JvdS provided key support to all the activities of the SAMI Galaxy Survey (`builder status'). All authors discussed the results and commented on the manuscript.  


\bibliographystyle{mnras}
\bibliography{biblioSAMI} 


\appendix

\section{Central velocity dispersions from GAMA spectra}
\label{Central velocity dispersions from GAMA spectra}

For a consistent comparison with some previous works, such as \citet{Bluck2022}, who use fibre
measurements of velocity dispersions in addition to those derived
from spatially-resolved spectroscopy, we take advantage of the central velocity dispersion ($\sigma_{\star}$) measured from the GAMA fibre spectra from DR4 \citep{Driver2022}. These measurements also allow a comparison with $\sigma_{\rm c}$ values from the SAMI Galaxy Survey (see Section~\ref{Central velocity dispersions}). Moreover, $\sigma_{\star}$ values enable us to extend the analysis of the roles of bulge mass and central velocity dispersion in star formation quenching to the whole GAMA equatorial survey regions up to $z<0.1$ (see Appendix~\ref{Role of bulge mass and central velocity dispersion in quenching for GAMA galaxies}).

In brief, $\sigma_{\star}$ is measured with the pPXF software using the best-fitting templates derived from the MILES library of stellar spectra. Due to the heterogeneity of the GAMA spectra (which incorporate various spectroscopic galaxy surveys), $\sigma_{\star}$ values have been calibrated and cross-validated via both intra- and inter-survey comparisons in order to quantify random and systematic errors in the measurements as a function of S/N, velocity dispersion, and survey. To select reliable estimates of $\sigma_{\star}$, we apply the following cuts: $5<\sigma_{\star}< 450$ km$\,\rm s^{-1}$ and $\sigma_{\star, \rm err}$< $\sigma_{\star}\times0.2+25$ km$\,\rm s^{-1}$, which are the criteria used by \citet{Dogruel2023} to reproduce the $M_{\star}$--$\sigma_{\star}$ relation in the GAMA survey.

\begin{figure}
\includegraphics[scale=0.29]{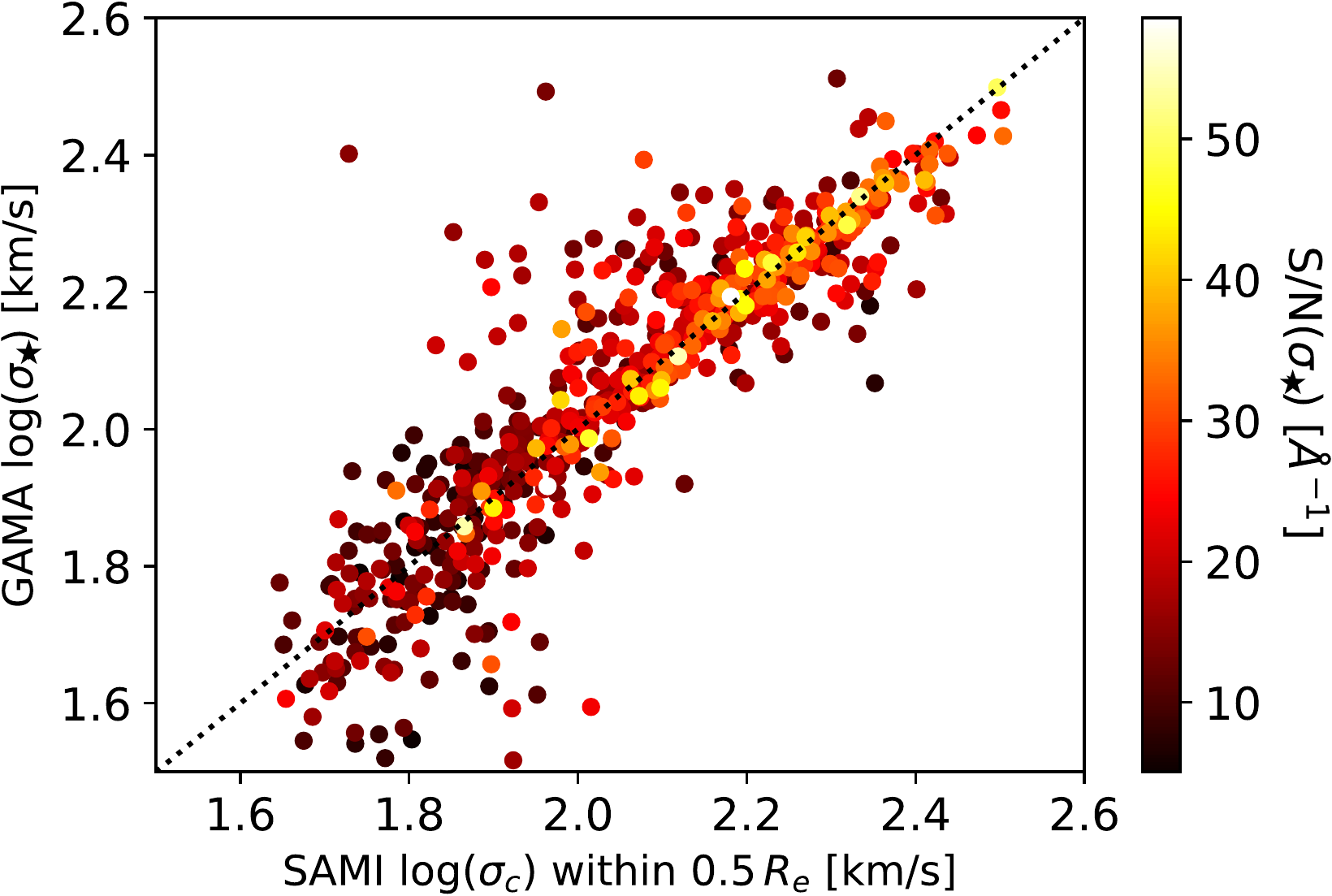}
\caption{Comparison of the central stellar velocity dispersion $\sigma_{\rm c}$ within $0.5\,R_{\rm e}$ and the stellar velocity dispersion $\sigma_{\star}$ from the GAMA fibre spectra, colour-coded by the signal-to-noise ratio for $\sigma_{\star}$. The dotted black line shows the 1:1 relation. }
\label{Sigma_diagrams}
\end{figure}

Out of the 977 SAMI galaxies that characterise Sample~A, 755 have reliable estimates of $\sigma_{\star}$. For this subset, Figure~\ref{Sigma_diagrams} shows the comparison between $\sigma_{\rm c}$ and $\sigma_{\star}$, colour-coded by the signal-to-noise ratio for the $\sigma_{\star}$ measurements. The estimates are grouped around the 1:1 relation, with a stronger agreement for higher values of S/N in $\sigma_{\star}$. The distribution of the differences between $\sigma_{\rm c}$ and $\sigma_{\star}$ has standard deviation SD $=0.11$. Only $\sim$1\% of galaxies have $\sigma_{\rm c}$  different from $\sigma_{\star}$ at $>3$\,SD. These outliers might be due to the fact that $\sigma_{\star}$ are mostly taken in 1.6\,arcsec (GAMA) or 3\,arcsec (SDSS) fibre apertures, which probe the $0.5\,R_{\rm e}$ aperture that characterises $\sigma_{\rm c}$ for $\sim$30\% of the SAMI galaxies and a more central region for the remaining SAMI galaxies. There is also the possibility that the GAMA fibres have not been placed exactly on the centroids of the galaxies. 

The results of this paper are consistent if we use $\sigma_{\rm c}$ or $\sigma_{\star}$. Moreover, our findings do not change if we use SAMI velocity dispersions estimated within a central 3\,arcsec aperture, in agreement with the comparison between this latter estimates and $\sigma_{\star}$ from \citet{Scott2017}.

\section{Examples of decision trees from Random Forest analysis}
\label{Examples of decision trees from Random Forest analysis}
\begin{figure*}
\includegraphics[scale=0.28]{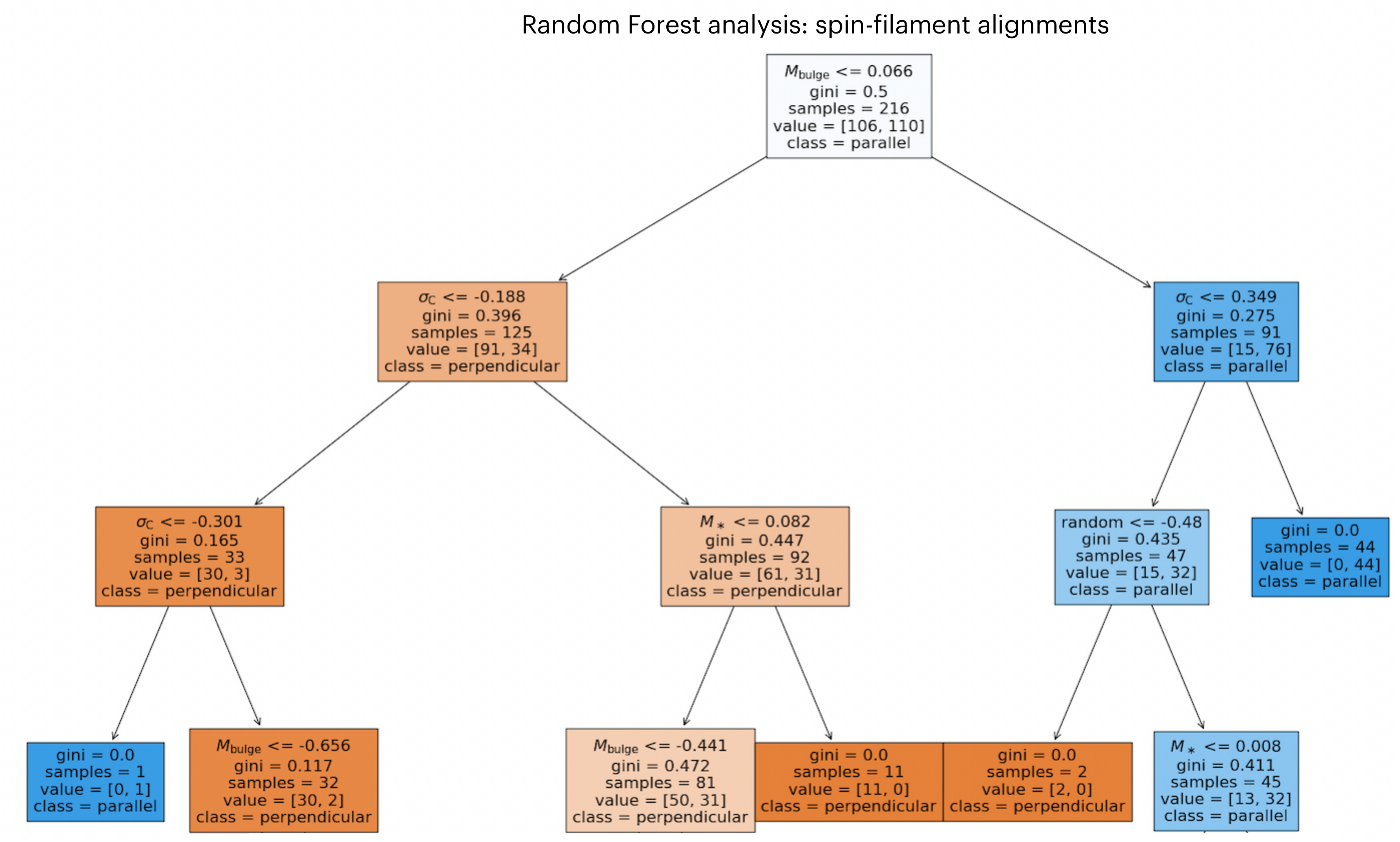}

\vspace{0.5cm}

\includegraphics[scale=0.28]{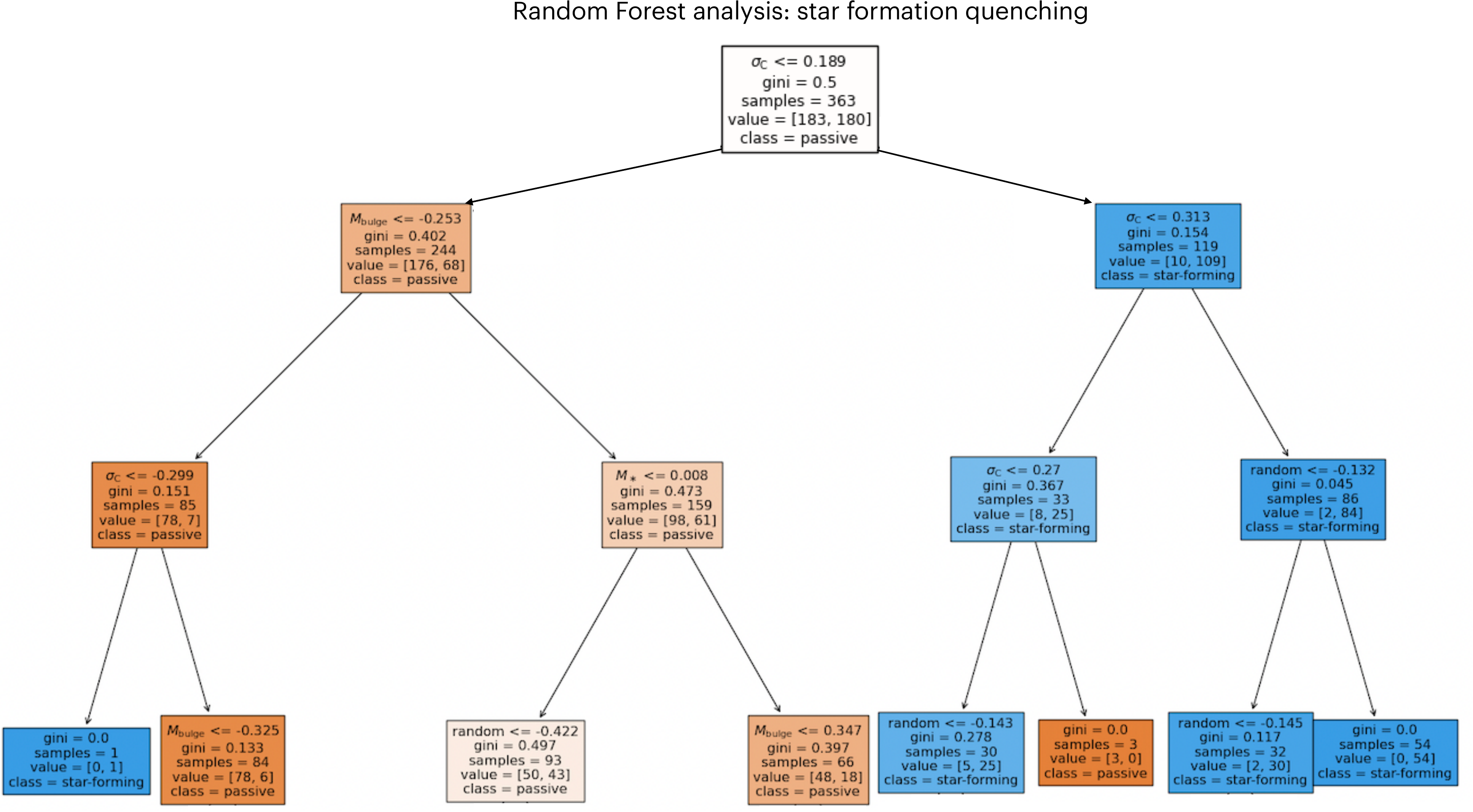}
\caption{Examples of two randomly selected decision trees out of the 150 implemented for training the Random Forest classifier. The top and bottom panels refer to the parallel versus perpendicular alignment classification and the
 star-forming versus passive classification, respectively. For both cases, we display three levels deep into the tree. For spin--filament alignments, $M_{\rm bulge}$ is selected as the first optimal parameter, while for star formation quenching $\sigma_{\rm c}$ is selected at the first split to minimise the Gini impurity. }
\label{RFtree}
\end{figure*}

 Figure~\ref{RFtree} shows examples of two randomly selected decision trees out of the 150 implemented for training the Random Forest classifier. The top and bottom panels refer to the parallel versus perpendicular alignment classification and the
 star-forming versus passive classification, respectively. For both cases, we display three levels deep into the tree. Each box presents the optimal parameter and its value for the classification, the Gini impurity and the sample size. The starting value of the Gini impurity at 0.5 is expected for matched samples, i.e. where the size of the two classes is constructed to be the same in order to avoid possible biases towards the largest population. The colour of the boxes becomes redder as the classifier identifies more passive galaxies or galaxies with a perpendicular alignment, while it becomes bluer for more star-forming galaxies or for galaxies with a parallel tendency. For spin--filament alignments, $M_{\rm bulge}$ is selected as the first optimal parameter, while for star formation quenching $\sigma_{\rm c}$ is selected at the first split to minimise the Gini impurity. As these galaxy parameters are the first ones chosen by the classifiers for more and more decision trees, they acquire the 
 highest relative importance with respect to the other explored properties in solving the classification problems.

\section{Role of bulge mass and central velocity dispersion in quenching for nearby GAMA galaxies}
\label{Role of bulge mass and central velocity dispersion in quenching for GAMA galaxies}

\begin{table}
\centering
\caption{Results from Spearman rank correlation test and PLS technique about star formation quenching for GAMA galaxies within $z<0.1$. Column~1 lists the analysed galaxy property, column~2 the number of galaxies, column~3 the Spearman correlation coefficient, column~4 the $p$-value from the Spearman test with significant $p$-values ($<$0.05) highlighted in bold, and column~5 the explained variance in $\Delta$(SFR).}
\label{SpearmanResultsGAMA}
\begin{tabular}{@{}lcccc@{}}
& &  $\Delta$(SFR) & & \\
\toprule
Galaxy property        & $N_{\rm gal}$ & $\rho$  & $p_{\rm S}$& Variance (\%)\\
\midrule
$\log(M_{\rm bulge}/M_{\odot})$ & 1415  & $-$0.40 & $\mathbf{10^{-37}}$ & 7.0 \\
B/T                             & 1415  & $-$0.39 & $\mathbf{10^{-36}}$ & 3.0  \\
$\log(M_\star/M_{\odot})$       & 1415  & $-$0.26 & $\mathbf{10^{-16}}$ & 1.2 \\
$\sigma_{\rm c}$                    & 1415  & $-$0.42 & $\mathbf{10^{-41}}$ & 89 \\
\bottomrule
\end{tabular}
\end{table}

\begin{table*}
\centering
\caption{Results from Random Forest classifier about star formation quenching for GAMA galaxies within $z<0.1$. Column~1 lists the studied process, column~2 the selected galaxy sub-sample, column~3 the number of galaxies, column~4 the classification, column~5 the Area Under the Curve values for the training sample and column~6 the {\fontfamily{qcr}\selectfont min\_samples\_leaf} parameter.}
\label{RFresultsGAMA}
\begin{tabular}{@{}lccccc@{}}
\toprule
Process & Selection & $N_{\rm gal}$ &  Classification ($N_{\rm gal}$) & AUC$_{\rm train}$  & {\fontfamily{qcr}\selectfont min\_samples\_leaf}\\
\midrule
$\Delta$(SFR) & All GAMA  & 1415 & PAS: $\Delta$(SFR) $<-1.0$ (433) &0.89 & 20 \\
& galaxies $z<0.1$ & & SF:\hspace{2mm} $\Delta$(SFR) $>-0.5$ (433)  \\
\\
$\Delta$(SFR) & GAMA centrals  & 939 &  PAS: $\Delta$(SFR) $<-1.0$ (200) &0.85 & 32 \\
& \& massive satellites & & SF:\hspace{2mm} $\Delta$(SFR) $>-0.5$ (200)  \\
\bottomrule
\end{tabular}
\end{table*}

We extend the study about the roles of bulge mass and central velocity dispersion in star formation quenching to the whole GAMA equatorial regions up to $z<0.1$, i.e.\ including those GAMA galaxies without SAMI data. For 1415 GAMA galaxies there are available $M_{\rm bulge}$ (see Section~\ref{Galaxy and bulge stellar masses}) and $\sigma_{\star}$ measurements (see Appendix~\ref{Central velocity dispersions from GAMA spectra}). The photometric SFRs are obtained from the spectral energy distribution (SED)-fitting code MAGPHYS \citep{daCunha2008,Driver2016}. Consistent results are found using spectroscopic SFR values measured following \citet{Barsanti2018}, where the H$\alpha$ luminosity is estimated from the procedure outlined in \citet{Hopkins2003} and \citet{Gunawardhana2013}. Like for the SAMI galaxies in Section~\ref{Star formation quenching}, star formation quenching is defined as the change in location from the main sequence in the stellar mass–star formation rate plane ($\Delta$ SFR).

We analyse the correlations between star formation quenching and bulge mass, bulge-to-total flux ratio, stellar mass and central velocity dispersion. Table~\ref{SpearmanResultsGAMA} reports the results from the Spearman rank correlation test and the PLS regression technique for the 1415 GAMA galaxies at $z<0.1$. The strongest correlation and the highest contribution to the variance in $\Delta$(SFR) is found for $\sigma_{\star}$. Controlling for the dependency on $\sigma_{\star}$, we do not find any significant residual correlations with the other parameters. On the other hand, a significant dependency with $\sigma_{\star}$ ($\rho=-0.15$, $p_{\rm S}=10^{-5}$) is detected for $\Delta$(SFR) while controlling for $M_{\rm bulge}$. These results are in agreement with our findings from Table~\ref{SpearmanResults} and the right panel of Figure~\ref{PartialCorrelations_SFR_Angle}.

\begin{figure}
\includegraphics[scale=0.27]{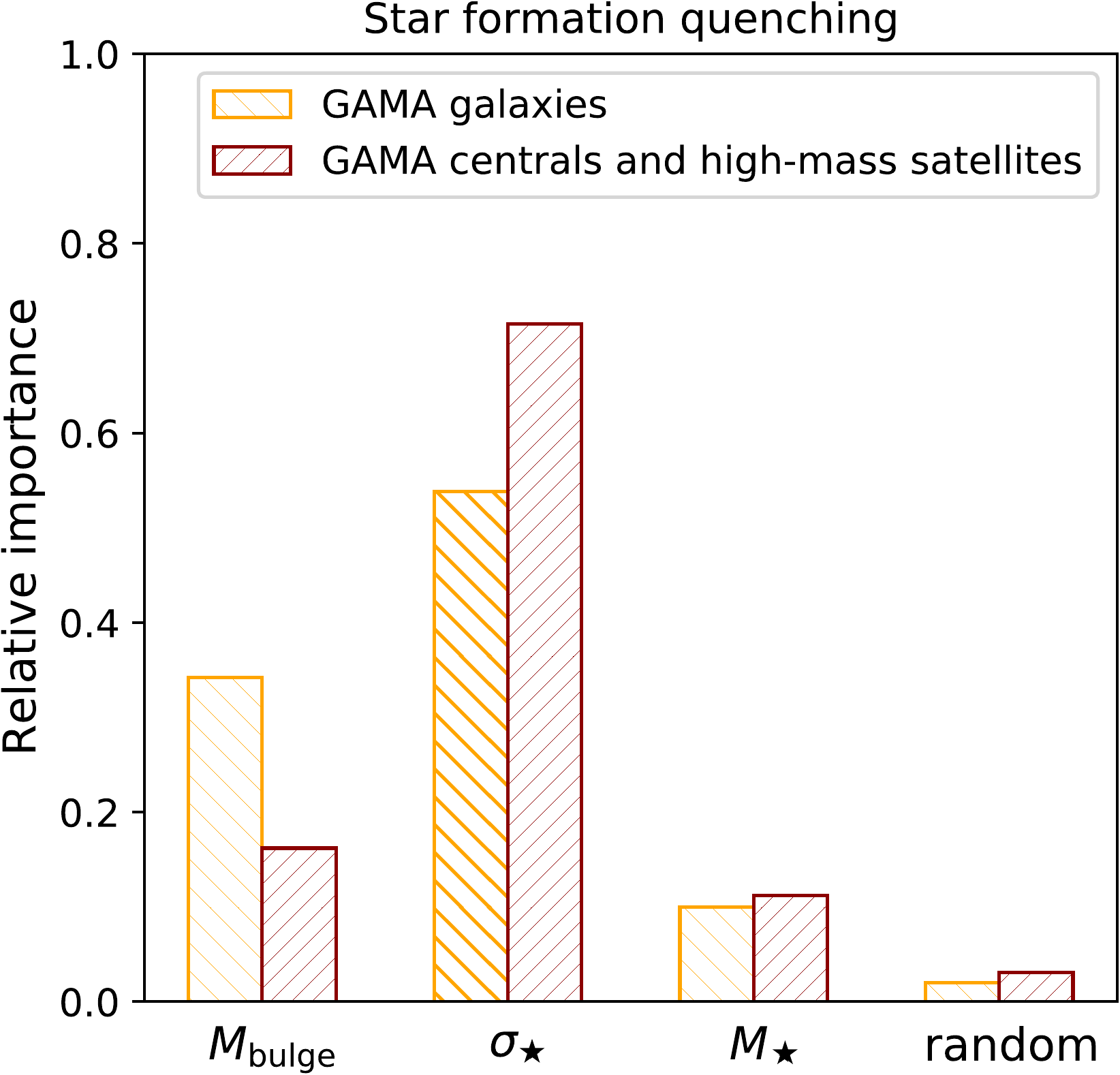}
\caption{Relative importance of each input galaxy parameter for star formation quenching for GAMA galaxies within $z<0.1$ (orange histogram) and for GAMA centrals and high-mass satellites (dark red histogram) from the Random Forest classifier. Central velocity dispersion is the most important parameter for separating passive and star-forming galaxies.}
\label{RFfigureGAMA}
\end{figure}

For the Random Forest classifier, we separate 788 passive galaxies with $\Delta$(SFR) $<-1.0$ from 433 star-forming galaxies with $\Delta$(SFR) $>-0.5$. Since the passive galaxy population is almost twice as large as the star-forming sample, in order to avoid for the algorithm to develop a bias towards the largest population, we sub-sample the passive sample to match the star forming population in size. Thus, the final sample is characterised by 433 passive galaxies and 433 star-forming galaxies. The algorithm performs well with AUC$_{\rm train}=0.89$ (see Table~\ref{RFresultsGAMA}) and the most predictive parameter is identified as the central velocity dispersion (see Figure~\ref{RFfigureGAMA}).

Since previous studies \citep{Bluck2020,Bluck2022,Piotrowska2022} found that it is the quenching of centrals and high-mass satellite galaxies that is primarily predicted by central velocity dispersion, out of the 1415 nearby GAMA galaxies we select 939 ($\sim66$\%) being centrals or satellites with $\log(M_\star/M_{\odot})>10$. We separate 385 passive from 200 star-forming, sub-sampling the passive population to match the star-forming size. The algorithm stabilises at AUC$_{\rm train}=0.85$ (see Table~\ref{RFresultsGAMA}) with the relative importance of $\sigma_{\star}$ increasing to 0.7 for the centrals and high-mass satellites (see Figure~\ref{RFfigureGAMA}) from 0.58 for the nearby GAMA galaxies. Overall, these outcomes for the GAMA survey are in agreement with our results from Section~\ref{Role of bulge mass and central velocity dispersion in spin--filament alignments versus quenching} for the SAMI Galaxy Survey and with the previous findings of \citet{Bluck2022} and \citet{Piotrowska2022} for the SDSS survey.

\bsp	
\label{lastpage}
\end{document}